\pdfoutput=1
\newcommand*{\ATLASLATEXPATH}{}
\documentclass[cernpreprint, texlive=2016, UKenglish]{\ATLASLATEXPATH atlasdoc}
 
\usepackage[subfig]{\ATLASLATEXPATH atlaspackage}
\usepackage[backref=false]{\ATLASLATEXPATH atlasbiblatex}

\usepackage[bsm,jetetmiss]{\ATLASLATEXPATH atlasphysics}
 
\usepackage{\ATLASLATEXPATH atlassnippets}
 
\graphicspath{{logos/}{common/figures/}}
 
\usepackage{ANA-SUSY-2018-17-PAPER-defs}
 
\usepackage{multirow}
 
\usepackage{todonotes}
 
\usepackage{xcolor}
\usepackage{dcolumn}

\PreprintIdNumber{CERN-EP-2020-123}
 
\AtlasTitle{Search for new phenomena in final states with large jet multiplicities and missing transverse momentum using $\sqrt{s} = 13$ TeV proton--proton collisions recorded by ATLAS in Run 2 of the LHC}
 
\AtlasAbstract{
Results of a search for new particles decaying into eight or more jets and moderate missing transverse momentum are presented. The analysis uses 139 $\mathrm{fb}^{-1}$ of proton--proton collision data at $\sqrt{s} = 13$ TeV collected by the ATLAS experiment at the Large Hadron Collider between 2015 and 2018. The selection rejects events containing isolated electrons or muons, and makes requirements according to the number of $b$-tagged jets and the scalar sum of masses of large-radius jets. The search extends previous analyses both in using a larger dataset and by employing improved jet and missing transverse momentum reconstruction methods which more cleanly separate signal from background processes. No evidence for physics beyond the Standard Model is found. The results are interpreted in the context of supersymmetry-inspired simplified models, significantly extending the limits on the gluino mass in those models. In particular, limits on the gluino mass are set at 2 TeV when the lightest neutralino is nearly massless in a model assuming a two-step cascade decay via the lightest chargino and second-lightest neutralino.
}
 
\author{The ATLAS Collaboration}
 
\AtlasRefCode{SUSY-2018-17}

\AtlasJournalRef{JHEP 10 (2020) 062}
\AtlasDOI{10.1007/JHEP10(2020)062}
 
\hypersetup{pdftitle={ATLAS document},pdfauthor={The ATLAS Collaboration}}
\usepackage{pifont}
 
\begin{document}
 
\maketitle

% The next lines are included from the .//source/intro.tex input file
\section{Introduction}
\label{sec:intro}

The Large Hadron Collider~\cite{Evans:2008zzb} (LHC) has produced a large dataset of proton--proton ($pp$) collisions at a centre-of-mass energy of 13~\TeV\, enabling searches for new heavy particles predicted by theories such as supersymmetry (SUSY)~\cite{Golfand:1971iw,Volkov:1973ix,Wess:1974tw,Wess:1974jb,Ferrara:1974pu,Salam:1974ig}. Evidence for SUSY models may be sought through searches for the production of these heavy particles (such as gluinos) decaying, often via extended cascades, into lighter ones. If the lightest of these interacts only weakly and is stable then it can be an ideal dark-matter candidate.

In $R$-parity-conserving (RPC)~\cite{Farrar:1978xj} SUSY models, the presence of a stable lightest supersymmetric particle (LSP) often leads to final states with significant missing transverse momentum (\met), often accompanied by a large number of jets. Large jet multiplicities would also occur in events in which gluinos decay via $R$-parity-violating (RPV)~\cite{Dreiner:1997uz} couplings on short ($\lesssim$\,ns) timescales. In this case the LSPs decay within the detector volume, so that the only invisible particles produced are neutrinos coming from SUSY particle decays, and hence the \met{} per event is generally smaller. A similar signature arises from any model in which cascade decays lead to the production of many jets, together with \met{} either from dark-matter particles or neutrinos.

This \paper reports the results of an analysis of \intlumi~\ifb\ of $pp$ collision data recorded at $\sqrt{s} = 13~\TeV$ by the ATLAS experiment~\cite{PERF-2007-01} throughout the entire Run-2 period of the LHC (2015--2018). It explores events with significant \met{} and at least eight jets with large transverse momentum (\pT).
Selected events are further classified into categories based on the presence of jets containing $b$-hadrons ($b$-jets) or on the sum of the masses of large-radius jets.
The $b$-jet selection improves sensitivity to beyond-the-Standard-Model (BSM) signals with enhanced heavy-flavour decays.
Given the unusually high jet multiplicities of the selected events, large jet masses can originate both from the capture of decay products from boosted heavy particles, including top quarks, and from accidental combinations~\cite{Cohen:2012yc}. The major backgrounds to the signal in this search are multijet production from QCD processes, top quark pair production (\ttbar) and $W$ boson production in association with jets (\Wjets).
 
Previous searches by ATLAS in similar final states were carried out on smaller LHC datasets recorded at $\sqrt{s}= 7~\TeV$ and $\sqrt{s} = 8~\TeV$ during 2011 and 2012~\cite{SUSY-2011-13,SUSY-2011-19,SUSY-2013-04}. In addition, two searches were performed at $\sqrt{s} = 13~\TeV$, one analysing the 2015 dataset~\cite{SUSY-2015-07} and one combining it with 2016 data~\cite{SUSY-2016-13} to achieve a total integrated luminosity of 36~\ifb.
The current analysis extends those previous studies by including the complete Run-2 LHC dataset, by using an optimised selection tailored to the increased integrated luminosity, and also by incorporating several improved analysis methods which further increase sensitivity. One such development is the use of the particle-flow jet and \met{} reconstruction algorithms recently developed for the ATLAS experiment~\cite{PERF-2015-09}. These algorithms combine measurements of inner tracker and calorimeter energy deposits to improve the accuracy of the charged-hadron measurement, leading to improvements in the jet and \met{} resolution and stability against additional $pp$ interactions in the same LHC bunch crossing. The analysis also employs an improved \met{} significance calculation~\cite{ATLAS-CONF-2018-038}, which accounts for the resolution of the reconstructed objects individually. This new definition increases the separation between events in which the \met{} originates from weakly interacting particles and those in which \met{} is only due to detector resolution effects.
The combination of the larger dataset and the developments in the analysis methodology provides this analysis with sensitivity over a significantly increased mass range.

% End of text imported from the .//source/intro.tex input file

% The next lines are included from the .//source/detector.tex input file
\section{ATLAS detector}
\label{sec:detector}

The ATLAS detector~\cite{PERF-2007-01} is a multipurpose particle detector with a nearly 4$\pi$ coverage in solid angle.\footnote{ATLAS uses a right-handed coordinate system with its origin at the nominal interaction point in the centre of the detector. The positive $x$-axis is defined by the direction from the interaction point to the centre of the LHC ring, with the positive $y$-axis pointing upwards, while the beam direction defines the $z$-axis. Cylindrical coordinates $(r,\phi)$ are used in the transverse plane, $\phi$ being the azimuthal angle around the $z$-axis. The pseudorapidity $\eta$ is defined in terms of the polar angle $\theta$ by $\eta=-\ln\tan(\theta/2)$. The rapidity is defined as $y$ = $0.5 \ln[(E + p_z)/(E-p_z)]$ where $E$ denotes the energy and $p_z$ is the component of the momentum along the beam direction. The transverse energy is defined to be $\ET=E \cos\theta$. The angular distance $\Delta R$ is defined as $\sqrt{(\Delta y)^2 + (\Delta \phi)^2}$.}  It consists of an inner tracking detector (ID) surrounded by a thin superconducting solenoid providing a \SI{2} {\tesla} axial magnetic field, electromagnetic (EM) and hadronic calorimeters, and a muon spectrometer. The inner tracking detector covers the pseudorapidity range \(|\eta| < 2.5\). It consists of silicon pixel, silicon microstrip, and transition radiation tracking detectors. A new inner pixel layer, the insertable B-layer~\cite{ATLAS-TDR-2010-19, PIX-2018-001}, was added at a mean radius of 3.3~cm before the start of the 2015 data-taking period, improving the identification of $b$-jets. Lead/liquid-argon (LAr) sampling calorimeters provide EM energy measurements with high granularity.
A steel/scintillator-tile hadron calorimeter covers the central pseudorapidity range (\(|\eta| < 1.7\)).
The endcap and forward regions are instrumented with LAr calorimeters for EM and hadronic energy measurements up to \(|\eta| = 4.9\).
The muon spectrometer surrounds the calorimeters and is based on three large air-core toroidal superconducting magnets with eight coils each.
The field integral of the toroids ranges between 2.0 and 6.0~Tm across most of the detector.
The muon spectrometer includes a system of precision tracking chambers and fast detectors for triggering.
A two-level trigger system \cite{TRIG-2016-01}~is used to select events.
The first-level trigger is implemented in hardware and uses a subset of the detector information to keep the accepted rate below \SI{100}{\kHz}.
This is followed by a software-based trigger that reduces the accepted event rate to \SI{1}{\kHz} on average.

% End of text imported from the .//source/detector.tex input file
 
% The next lines are included from the .//source/datasets.tex input file
\section{Datasets}
\label{sec:datasets}

\subsection{Data}
The events considered in this \paper correspond to \intlumi~\ifb\ of $pp$
LHC collision data collected between 2015 and 2018 by the ATLAS detector, at a centre-of-mass energy of 13\,TeV and with a 25\,ns proton bunch crossing interval.
In 2015--2016 the average number of interactions per bunch crossing (pile-up)
was $\langle\mu\rangle = 20$, increasing to $\langle\mu\rangle = 38$ in 2017 and $\langle\mu\rangle = 37$ in 2018.
The uncertainty in the combined 2015--2018 integrated luminosity is $1.7\%$~\cite{ATLAS-CONF-2019-021},
obtained using the LUCID-2 detector~\cite{Avoni:2018iuv} for the primary luminosity measurements.
 
Events were recorded using a variety of trigger selections. During both 2015 and 2016, events
were selected by a trigger requiring at least six jets with $\ET>45$\,\GeV{}
and central pseudorapidity, $|\eta|<2.4$. Further events were triggered in 2015
by requiring the presence of at least five jets with  $\ET>70$\,\GeV{} and $|\eta|<3.2$,
and in 2016 with a trigger requiring at least five jets with $\ET>65$\,\GeV{}
and $|\eta|<2.4$. In both 2017 and 2018, events were selected by triggers
requiring at least five jets with $\ET>70$\,\GeV{} and $|\eta|<2.4$,
or seven jets with $\ET>45$\,\GeV{} and $|\eta|<2.4$.
Additional triggers were also used in 2017 and for some periods in 2018, which required
at least six jets with $\ET>45$\,\GeV{} and $|\eta|<2.4$.
Due to their large trigger rates, these were set to select only a fraction of the events, approximately 15\% of the two years' data; such triggers are referred to as `prescaled' triggers.
The trigger jet calibration was improved in 2017--2018, resulting in
substantially improved efficiency~\cite{ATL-DAQ-PUB-2018-002} after offline selection for the same trigger thresholds.
The trigger selections are summarised later in \Tab{\ref{tab:SRdefs}} together with further event selections.
 
Data quality requirements are imposed to ensure that only events
in which the entire ATLAS detector was functioning well are used~\cite{DAPR-2018-01}. These, for example,
exclude events containing data corruption in the ID and calorimeters, excessive noise and spurious
jets produced by non-collision backgrounds~\cite{DAPR-2012-01,ATLAS-CONF-2015-029}.
 
\subsection{Monte Carlo simulations}
 
Simulated events produced with several Monte Carlo (MC) event generators
are used to predict yields for subdominant background contributions from Standard Model (SM) processes
and for possible signals.
 
All simulated events are overlaid with multiple $pp$ collisions simulated with the
soft QCD processes of \pythia~8.186~\cite{Sjostrand:2007gs} using the A3 set of tuned
parameters (A3 tune)~\cite{ATL-PHYS-PUB-2016-017} and the NNPDF2.3 LO
parton distribution functions (PDFs)~\cite{Ball:2012cx}. The simulated events are
required to pass the trigger selections, and are weighted such that the pile-up conditions match
those of the data. The response of the detector to particles was modelled with an ATLAS
detector simulation~\cite{SOFT-2010-01} based on \geant~\cite{Agostinelli:2002hh} (full simulation),
or using fast simulation based on a parameterisation of the performance of the
ATLAS EM and hadronic calorimeters~\cite{ATL-SOFT-PUB-2014-001} and
on \geant{} elsewhere.

\label{sec:dataMC:background}
 
For the generation of \ttbar and single top quarks via the $Wt$ process and in the $s$-channel,
matrix elements were calculated at next-to-leading order (NLO)
using the \powhegbox~v2~ 
generator~\cite{Nason:2004rx,Frixione:2007vw,Alioli:2010xd,Alioli:2009je,Re:2010bp,Alioli:2011as}
with the NNPDF3.0 NLO PDF set~\cite{Ball:2014uwa} in the five-flavour scheme.
Electroweak $t$-channel single-top-quark events were generated using \powhegbox~v2,
using the four-flavour scheme for the NLO
matrix element calculations together with the fixed four-flavour PDF set NNPDF3.04f NLO. The diagram removal scheme~\cite{Frixione:2008yi}
was used to prevent $Wt$ events from being counted as $\ttbar$ events beyond leading order (LO).
For these processes, the top quarks were decayed using \madspin~\cite{Frixione:2007zp,Artoisenet:2012st} preserving
all spin correlations, while for all processes the parton shower, fragmentation,
and the underlying event were simulated using \pythia~8.230~\cite{Sjostrand:2014zea}
with the NNPDF2.3 LO PDF set and the ATLAS A14 tune~\cite{ATL-PHYS-PUB-2014-021}.
The top quark mass was set to $172.5$\,\GeV. The $h_{\mathrm{damp}}$ parameter, which controls
the $\pt$ of the first additional emission beyond the Born configuration in \powheg, was set to $1.5$
times the mass of the top quark.
The main effect of this parameter
is to regulate the high-$\pt$ emission against which the \ttbar system recoils.
The \evtgen~v1.6.0 program~\cite{Lange:2001uf} was
used to model properties of the $b$- and $c$-hadron decays for this process
and all others not simulated with \sherpa~\cite{Gleisberg:2008ta} unless otherwise stated.
Simulated \ttbar events are normalised to the cross-section calculated at
next-to-next-to-leading order (NNLO) in perturbative QCD, including soft-gluon
resummation to next-to-next-to-leading-logarithm (NNLL) accuracy~\cite{top++}.
The single-top-quark events for the $Wt$ channel are normalised using its approximate NNLO prediction~\cite{Kidonakis:2010ux,Kidonakis:2013zqa}, while the $t$- and $s$-channels are normalised using their NLO predictions~\cite{Aliev:2010zk,Kant:2014oha}.
 
Events containing \ttbar and additional heavy particles -- comprising three or four top quarks, \ttW, \ttZ and
$\ttbar + WW$ production -- were modelled using \MGMCatNLOV~\cite{Alwall:2014hca} for the matrix element calculation,
interfaced to the \pythia8 parton shower, hadronisation and underlying event model.
The $\ttbar + WW$, three- and four-top-quark processes were simulated at LO
in the strong coupling constant $\alphas$, using \MGMCatNLOV{v2.2.2} interfaced to \pythia~8.186.
The predicted production cross-sections were calculated to NLO as described in Ref.~\cite{Alwall:2014hca}
for these processes other than three-top-quark production, for which the cross-section was calculated to LO.
The production of \ttW and \ttZ events was modelled using
\MGMCatNLOV{v2.3.3} at NLO with the NNPDF3.0 NLO PDF. Top quarks were decayed
at LO using \madspin{} to
preserve spin correlations. Parton shower and hadronisation were modelled with
\pythia~8.210.
The cross-sections were calculated at NLO QCD and NLO EW accuracy using
\amcatnlo as reported in Ref.~\cite{deFlorian:2016spz}.
In the case of \ttll the cross-section is additionally scaled by
an off-shell correction estimated at one-loop level in \alphas.
For all processes, the A14 set of \pythia 8 parameters
was used, together with the NNPDF2.3 LO PDF set.
\evtgen~v1.2.0 was used to model properties of the $b$- and $c$-hadron decays.
The contribution from $\ttbar + H$ was checked and found to be negligible.
 
Events containing $W$ or $Z$ bosons associated with jets were simulated using the \sherpa~v2.2.1 generator.
Matrix elements were calculated for up to two partons at NLO and four partons at LO using the
\comix~\cite{Gleisberg:2008fv} and \openloops~\cite{Cascioli:2011va} matrix element generators and merged with the
\sherpa parton shower~\cite{Schumann:2007mg} using the ME+PS@NLO prescription~\cite{Hoeche:2009rj}.
The NNPDF3.0 NNLO PDF set~\cite{Ball:2014uwa}
was used in association with a tuning performed by the \sherpa authors.
 
Diboson processes with one hadronically decaying boson accompanied by one charged lepton and one neutrino,
two charged leptons or two neutrinos were simulated using \sherpa~v2.1.1.
The calculations include one additional parton at NLO for $ZZ\to 2\ell+q\bar{q}$ and $ZZ\to 2\nu+q\bar{q}$ only,
and up to three additional partons at LO using the \comix and \openloops matrix element generators and merged with
the \sherpa parton shower using the ME+PS@NLO prescription.
The NNPDF3.0 NNLO PDF set was used in conjunction with a dedicated parton shower tuning
developed by the \sherpa authors.
Diboson processes with four charged leptons, three charged leptons and one neutrino,
or two charged leptons and two neutrinos, are found to be negligible.
 
Theoretical uncertainties are considered for all simulated samples.
By far the most important process simulated in this analysis is \ttbar production,
and several samples produced with different configurations, as explained below, are compared to estimate the uncertainty in this background.
Samples were produced with the factorisation and renormalisation scales varied coherently up and down by a factor of two,
and with parameters set to provide more/less radiation in the parton shower~\cite{ATL-PHYS-PUB-2018-009}.
Additionally, to account for uncertainties from the parton shower modelling and generator choice,
the nominal sample is compared to a sample generated with
\powhegbox interfaced to \herwig7~\cite{Bahr:2008pv} using the H7-UE tune~\cite{Bellm:2015jjp}
and the MMHT2014 LO PDF set~\cite{Harland_Lang_2015}, as well as samples generated with
\MGMCatNLO interfaced to \pythia8. These alternative samples each use the NNPDF3.0 NLO PDF for the matrix element calculation.
The comparison with samples which vary the amount of additional radiation
contributes the largest uncertainty in the \ttbar signal region predictions.
Similar alternative samples are used to assess the uncertainties in single-top-quark production, whereas uncertainties in other processes are handled via scale variations in the corresponding generator.
 
Full simulation was used for all background MC samples,
ensuring an accurate representation of detector effects.
Further details of samples can be found in
Refs.~\cite{ATL-PHYS-PUB-2017-005,ATL-PHYS-PUB-2017-006,ATL-PHYS-PUB-2018-009}.
 
\label{sec:dataMC:signal}
 
A number of SUSY signal model samples were simulated using the ATLAS fast detector simulation~\cite{SOFT-2010-01} to allow the
interpretation of the search results in terms of SUSY parameters.
Substantial cross-sections are possible for production of gluinos. The resulting cascade decays result in a large multiplicity
of jets, and may also exhibit an unusually high heavy-flavour content or
atypically large jet masses.
 
The first type of SUSY signal simulated is a simplified model, in which gluinos are pair-produced and
then decay through an off-shell squark via the cascade:
\begin{align*}
\gluino   & \to q + \bar{q}' + \chinoonepm \hspace*{0.3cm} \left(q,q^\prime \in \{ u,d,s,c\}\right), \\
\chinoonepm & \to W^{\pm} + \ninotwo, \\
\ninotwo  & \to Z + \ninoone,
\end{align*}
where the quarks are only permitted to be from the first two generations.
The parameters of the model are the masses of the gluino, $m_{\gluino}$,
and of the lightest neutralino, $m_{\ninoone}$.
The mass of the $\chinoonepm$  is constrained to be $(m_{\gluino} + m_{\ninoone})/2$,
and the mass of the $\ninotwo$ is set to $(m_{\chinoonepm} + m_{\ninoone})/2$.
A diagram of this `two-step' simplified model is shown in Figure~\ref{samples:fig:feynman_diagrams}(a).

An additional signal model to which this analysis has significant sensitivity is
gluino-mediated top squark ($\stopone$) production, in which top-quark-rich final states are produced as shown in Figure~\ref{samples:fig:feynman_diagrams}(b).
This model manifests itself as top quark production via either off-shell or on-shell top squarks.
In the off-shell model, pair-production of gluinos is followed by their decay with a 100\% branching ratio into $\ttbar + \ninoone$,
through a virtual top squark. Naturalness arguments for SUSY favour light gluinos,
top squarks, and Higgsinos, so they motivate consideration of this final state.

\begin{figure}
\centering
\begin{tabular}{cc}
\subfloat[Two-step decay]{\includegraphics[width=0.31\textwidth]{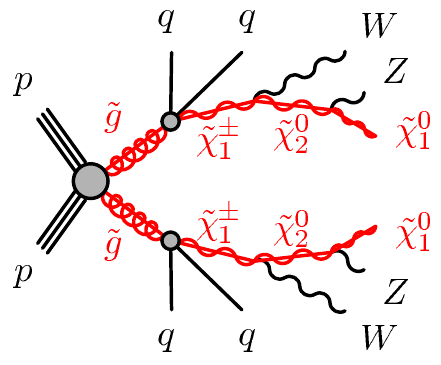}}\hspace{2ex}
\subfloat[Off-shell top squarks]{\includegraphics[width=0.31\textwidth]{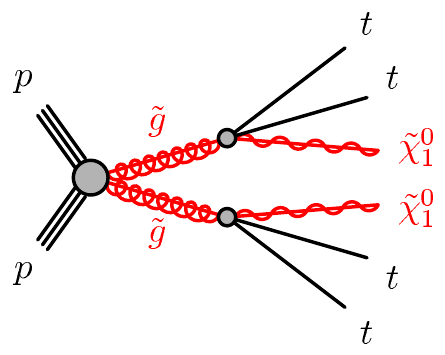}}\hspace{2ex}
\subfloat[RPV]{\includegraphics[width=0.31\textwidth]{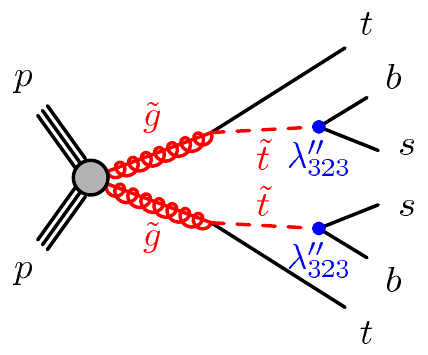}}
\end{tabular}
\caption{Pseudo-Feynman diagrams for the different signal models used in this search. In (c), $\lambda''_{323}$ is one of the couplings of the third-generation squark to quarks in the RPV model. }
\label{samples:fig:feynman_diagrams}
\end{figure}

Permitting non-zero RPV couplings allows consideration of another variety of
gluino-mediated top squark production, wherein the last step of the decay proceeds via a
baryon-number-violating interaction: $\stopone \rightarrow \bar{s} + \bar{b}$ (with charge conjugates
implied). Such RPV models may give rise to final states with missing transverse momentum,
for example from leptonic decays of $W$ bosons produced in top quark decays.
The current analysis accepts final states with sufficiently low missing transverse momentum to be sensitive to these RPV scenarios.
Figure~\ref{samples:fig:feynman_diagrams}(c) presents the RPV simplified model considered, for which the coupling strength induces prompt top squark decays.

The signal samples were generated using \MGMCatNLO interfaced to \pythia8 with the A14 tune
for the modelling of the parton shower, hadronisation and underlying event.
The versions of the generators used for
the two-step and RPV simplified models were \MGMCatNLOV{v2.6.2} with \pythia~8.212,
and for the gluino-mediated top squark model \MGMCatNLOV{v2.3.3} with \pythia~8.212.
The matrix element calculation was performed at tree level and includes the emission of up to two additional partons.
The PDF set used for the generation was NNPDF2.3 LO. The \evtgen~v1.6.0 program
was used to simulate properties of the $b$- and $c$-hadron decays.
The matrix-element to parton-shower matching was done using the CKKW-L prescription~\cite{Lonnblad:2001iq},
with a matching scale set to $m_{\gluino}/4$.

Signal cross-sections were calculated to approximate NNLO in the strong coupling constant, adding the resummation of soft gluon emission at NNLL accuracy~\cite{Beenakker:2016lwe,Beenakker:2014sma,Beenakker:2013mva,Beenakker:2011sf,Beenakker:2009ha,Kulesza:2009kq,Kulesza:2008jb,Beenakker:1996ch}.
The nominal cross-section and its uncertainty were determined using the PDF4LHC15\_mc PDF set, following the recommendations of Ref.~\cite{Butterworth:2015oua}.
 
% End of text imported from the .//source/datasets.tex input file
 
% The next lines are included from the .//source/objevtsel.tex input file
\label{sec:objevtdef}

\section{Reconstruction and particle identification}
\label{sec:objdef}
 
Primary vertices are reconstructed using at least two charged-particle tracks with $\pT > 500$~\MeV{} measured
by the ID~\cite{ATL-PHYS-PUB-2015-026}. The primary vertex with the largest sum of squared track transverse momenta ($\sum{\pT^2}$)  is
designated as the hard-scatter vertex.
 
Jets are reconstructed using the anti-$k_t$~\cite{Cacciari:2008gp,Fastjet,PERF-2014-07}~jet algorithm with radius parameter $R=0.4$.
It uses as inputs particle-flow objects. These are charged-particle tracks matched to the hard-scatter vertex with the requirement $|z_0 \sin{\theta}|<2.0~$mm, where $z_0$ is the longitudinal impact parameter, and calorimeter energy clusters surviving an energy subtraction algorithm that removes the calorimeter deposits of good-quality tracks originating from any vertex~\cite{PERF-2015-09}.
To eliminate jets containing a large energy contribution from pile-up, jets are tested for compatibility with the hard-scatter vertex with the jet vertex tagger (JVT) discriminant, utilising information from the ID tracks associated with the jet~\cite{PERF-2014-03}.
Any jets with $20~\GeV{} < \pT < 120$~\GeV{} and $|\eta|<2.4$ for which $\text{JVT} < 0.5$ are considered to originate from pile-up and are therefore rejected from the analysis.
After the selection above is applied, only jets with $\pT > 20$~\GeV{} and $|\eta|<2.8$ are considered in this analysis, with the exception of the \MET{} calculation, for which jets in the range $2.8\leq |\eta| \leq 4.5$ are also used. Hadronically decaying $\tau$-leptons are not discriminated from other hadronic jets.

Reconstructed $R=0.4$ jets with $\pT > 20$~\GeV{} and $|\eta|<2.0$ are reclustered to form large-radius jets using the anti-$k_t$ algorithm with radius parameter $R=1.0$~\cite{ATLAS-CONF-2017-062}.
The input jets are required to pass an overlap removal procedure accounting for ambiguities between jets and leptons, as
discussed below. Large-radius jets are retained for analysis if they have $\pT > 100~$\GeV{} and $|\eta|<1.5$.

Jets containing $b$-hadrons and which are within the ID
acceptance ($|\eta|<2.5$) are identified as $b$-tagged jets using a multivariate algorithm that exploits the impact parameters of the charged-particle tracks, the presence of secondary vertices, and the reconstructed flight paths of $b$- and $c$-hadrons inside
the jet~\cite{ATL-PHYS-PUB-2016-012}.
The output of the algorithm is a single discriminant value which signifies the likelihood that the jet contains a $b$-hadron.
This analysis considers jets to be $b$-tagged if the discriminant exceeds a threshold that results in an average identification efficiency of $70\%$ for jets containing $b$-hadrons in simulated \ttbar\ events~\cite{FTAG-2018-01}.
In the same event sample, a rejection factor of approximately 300 is reached for jets initiated by light quarks and gluons and 8.9 for jets initiated by charm quarks.

Electron candidates are reconstructed from energy deposits in the EM calorimeter that are matched to charged-particle tracks in the ID~\cite{EGAM-2018-01}.
`Baseline electrons' are required to satisfy $\pt>7$~\GeV\ and $|\eta|<2.47$. They are identified using the `loose' operating point provided by a likelihood-based algorithm~\cite{EGAM-2018-01}.
Electrons with $\pT>20~\GeV$ are defined as `signal electrons' if they pass a `tight' likelihood selection including impact parameter restrictions and the `GradientLoose' isolation requirement~\cite{PERF-2016-01} in addition to the `baseline electron' preselection.
To achieve additional rejection of background electrons from non-prompt sources, signal electron tracks must be matched to the hard-scatter vertex with a longitudinal impact parameter $|z_0\sin\theta|<0.5$~mm and a transverse impact parameter significance $|d_0|/\sigma(d_0)<5$. Signal electrons are used in leptonic control regions, as described in \Sect{\ref{sec:bkgest:leptonic}}. The electron reconstruction and identification efficiency in simulated samples are corrected by factors determined by data--MC comparison using a given final state~\cite{PERF-2016-01}.

Photon candidates are identified using `tight' criteria for lateral shower shapes in the first and second layers of the EM calorimeter, as well as for the degree of hadronic shower leakage~\cite{EGAM-2018-01}.
Acceptance requirements of $\pT>40~\GeV$ and $|\eta|<2.37$ are applied.
Additionally, photons falling in the region $1.37<|\eta|<1.52$ are removed, to avoid a region of the calorimeter with limited instrumentation.

Muon candidates are reconstructed from matching tracks in the ID and muon spectrometer, refined through a global fit which uses the hits from both subdetectors~\cite{PERF-2015-10}.
`Baseline muons' must have $\pt>6$~\GeV\ and $|\eta|<2.7$, and satisfy the `medium' identification criteria.
Similarly to electrons, the longitudinal impact parameter $z_0$ relative to the hard-scatter vertex is required to satisfy $|z_0\sin\theta| <0.5$~mm.
Muons are characterised as `signal muons' if they have a higher transverse momentum, $\pT>20~\GeV$, and satisfy the `GradientLoose' isolation requirement~\cite{PERF-2015-10}, as well as a further transverse impact parameter restriction $|d_0|/\sigma(d_0)<3$. Signal muons are used in leptonic control regions, as described in \Sect{\ref{sec:bkgest:leptonic}}.
Muon reconstruction and identification efficiencies in simulated samples are corrected with factors evaluated by a data--MC comparison~\cite{ATL-PHYS-PUB-2015-037}.

To resolve the reconstruction ambiguities between electrons, muons, photons and jets, an overlap removal procedure is applied to baseline objects.
First, any electron sharing an ID track with a muon is rejected. If it shares the same ID track with another electron, the one with lower \pt is discarded. Next, photons with $\Delta R < 0.4$ relative to an electron or a muon are discarded.
Subsequently, non-$b$-tagged jets are rejected if they lie within  $\Delta R = 0.2$ of an electron or if the jet has no more than three tracks with $\pT>500~\MeV$, or contains an ID track matched to a muon such that $\ptjet<2\pt^\mu$ and the muon track has more than 70\% of the sum of the transverse momenta of all tracks in the jet, such that the jet resembles radiation from the muon.
Finally, electrons or muons with $\Delta R < 0.4$ from a surviving jet are eliminated and jets with $\Delta R < 0.4$ from photons are removed.

The missing transverse momentum, \met, is defined as the magnitude of the negative vector sum of the transverse momenta of baseline electrons and muons, photons and jets, which pass an overlap removal procedure, based on removing duplicated energy contributions and therefore distinct from that used for jet/lepton disambiguation. A `soft term' is added to recover the contributions from other low-\pT{} particles~\cite{PERF-2016-07,ATLAS-CONF-2018-023}.
The soft term is constructed from all tracks that are not associated with any of the preceding objects, but that are associated with the hard-scatter vertex.
In this way, the missing transverse momentum is adjusted for the best calibration of the leptons, photons and jets, while maintaining pile-up robustness in the soft term.

The \met\ significance \metsig\ is defined in order to test the hypothesis that the total transverse  momentum carried by non-interacting particles 
is equal to zero against the hypothesis that
it is different from zero. It quantifies the degree to which the measured \met\ is inconsistent with originating purely from fluctuations and mismeasurements.
A large value indicates that the \met\ genuinely originates from non-interacting particles.
In this search, the variable \metsig characterises the \met according to the \pT, \pT resolution, and $\phi$ resolution of all  objects in the event and is defined~\cite{ATLAS-CONF-2018-038} as:
\begin{equation*}
\mathcal{S}(\met) ={ \frac{\met}{\sqrt{\sigma_\textrm{L}^2(1-\rho_\textrm{LT}^2)}}}.
\label{eq:objmetsig}
\end{equation*}
In this definition, $\sigma_\textrm{L}$ is the total expected longitudinal resolution of all objects in the event as a function of the \pT of each object. Likewise, $\rho_\textrm{LT}$ is the correlation factor between all longitudinal and transverse object resolutions.
The resolution of the soft term is fixed to $8.9~\GeV$ in each of the transverse coordinates.
The resolution of each jet is further modified by the probability that the jet is a pile-up jet mistakenly tagged as being from the hard-scatter interaction, parameterised in $\pt$, $\eta$ and JVT.
By individually accounting for these resolutions, this `object-based' \metsig\ definition captures the response of the ATLAS detector and reconstruction algorithms better than the simpler \metsig\ proxies used in previous searches (such as in Ref.~\cite{SUSY-2016-13}).

\section{Event selection}
 
Signal events for this analysis are characterised by a large jet multiplicity, beyond that generated by high cross-section SM processes, combined with an \met value that is significantly larger than that expected purely from detector resolution and mismeasurement effects.
Several signal regions are defined that select a minimum jet multiplicity and further require a large value of \metsig.

\label{sec:evtsel:srdef}
 
The common selection of events for the signal regions is as follows.
Events are rejected if no primary vertex is reconstructed.
To limit the contribution from SM background processes in which neutrinos are produced via the decay $\Wboson\to\ell\nu$,
leading to significant \met, events containing any baseline electron or muon with $\pT>10~\GeV$ surviving the overlap removal are rejected.

Biases in the \met\ due to pile-up jets surviving the JVT selection are removed by excluding events for which a jet $j$ with $60~\GeV{} <\pT<70~\GeV$ and $\mathrm{JVT}<0.2$ lies opposite to the \met\ vector in $\phi$ ($\left|\Delta\phi(j,\MET)\right| > 2.2$).
Likewise, events are rejected if they contain a jet with $\pT>50~\GeV$ and $|\eta|<2.0$ pointing in $\eta$--$\phi$ towards a region in which tile calorimeter modules were disabled.
 
Subsequently, restrictions on the jet multiplicity \njet\ are imposed; only jets with $\pT>50 ~(80)~\GeV$ and $|\eta|<2.0$ are considered as \emph{signal jets} and
therefore used in the \njet selection.
These selections are abbreviated as {\texttt j50} ({\texttt j80}), for which the corresponding jet multiplicities are denoted \njfifty (\njeighty).
The lower and higher jet-\pt thresholds provide sensitivity to a variety of potential SUSY mass spectra and ensure that the corresponding trigger selections give efficiencies in simulation and in data that match to within 2\%.
 
A requirement of $\metsig>5$ is the last element of the common selection.
This criterion eliminates the vast majority of SM multijet and other background events with low \met, while retaining sensitivity to a broad range of potential signals.

Motivated by the likelihood of heavy-flavour jets being produced from cascade decays, events are classified by their $b$-jet multiplicity.
Events passing the common selection may then be categorised according to three classes depending on whether they satisfy $\nbjet =0$, $\nbjet =1$ or $\nbjet \ge 2$.

Should SUSY particles be produced and decay through a long decay chain, or provide enough kinetic energy to significantly boost heavy particles such as top quarks or SM electroweak bosons, signal events might be characterised not only by an unusually large jet multiplicity but also by the formation of large-radius jets with high masses.
The kinematic structure of SM events, by contrast, does not produce a high rate of events containing large-radius jets with a mass greater than the top quark mass.
On this basis, the selection variable \mjsigma is defined to be the sum of the masses $m_j^{R=1.0}$ of the large-radius jets
\begin{equation*}
\mjsigma = \sum_j m_j^{R=1.0}
\end{equation*}
where the sum is over the large-radius jets that satisfy ${\pt}_{j}^{R=1.0} > 100$~\GeV\ and $|{\eta}_{j}^{R=1.0}| < 1.5$, as described in \Sect{\ref{sec:objdef}}.
After the common selection and \nbjet classification, selected events may be further partitioned into three \mjsigma bins:
\begin{itemize}
\item $\mjsigma \le 340~\GeV$;
\item $340~\GeV < \mjsigma \le 500~\GeV$;
\item $\mjsigma > 500~\GeV$.
\end{itemize}
As the bin boundaries are at approximately twice and three times the top quark mass, the residual irreducible backgrounds remaining in the higher \mjsigma bins are respectively top quark pair production in association with vector bosons and processes with four top quarks, both of which have a very small rate.

The subselections in \nbjet and \mjsigma are used in two subanalyses.
First, a set of multi-bin signal selections is defined to improve the exclusion reach in the absence of an observed excess.
Starting from the common selections, shown in \Tab{\ref{tab:SRdefs}}, for $\njfifty\ge 8$,
the selected events are partitioned by \nbjet and \mjsigma according to the thresholds above, defining nine statistically independent bins, as shown in \Tab{\ref{tab:SRVRSummary}}.
The same subdivision is applied for events satisfying the $\njfifty \ge 9$ and $\njfifty \ge 10$ common selections.
The statistical analyses of the three different $\njfifty$ selections are performed separately
since the selections are not statistically independent. For example, one of the nine bins in the 8-jet selection (e.g. $\nbjet=0$, $\mjsigma>500$) includes all events in the corresponding bin for the 9- and 10-jet selections.
The inputs to the statistical analysis also include validation regions (VRs) and control regions (CRs), further discussed in \Sect{\ref{sec:bkgest}}.
 
Additionally, an ensemble of single-bin regions is defined making use of inclusive selections in \nbjet and \mjsigma, which are designed to reject the SM hypothesis with high confidence, should new physics be present.
Following an optimisation procedure to maximise coverage over the benchmark signal models, the signal regions defined in \Tab{\ref{tab:SRdefsSingleBins}} were considered.
To give an indication of their relevance, the two-step gluino cascade decay motivates the \texttt{SR-10ij50-0ib-MJ340} and \texttt{SR-10ij50-0ib-MJ500} regions 
while the \texttt{SR-9ij50-0ib-MJ340}
and \texttt{SR-10ij50-1ib-MJ500}
regions are optimised for sensitivity to gluino decays into $\ttbar + \ninoone$.
The \texttt{SR-10ij50-1ib-MJ500} selection also provides the best sensitivity to RPV gluino decays.
 
If an excess of data is observed over the background predictions, its significance can be quantified in the single-bin regions.
If on the contrary the data are consistent with the background-only hypothesis, signal models are excluded using the constraint achieved by whichever signal region is expected to be most sensitive to the model, whether it is a single- or multi-bin selection.

\begin{table}
\renewcommand\arraystretch{1.4}
\centering
\begin{tabular}{c|p{5cm} p{5cm}}
\toprule
Selection criterion & \multicolumn{2}{c}{Selection ranges} \\
\midrule
Jet multiplicity, \njet
& \centering $\njfifty \ge \big\{8, 9, 10, 11, 12\big\}$  & \centering $\njeighty \ge 9$ \tabularnewline \hline
\multirow{3}{*}{Trigger thresholds}
& \multirow{2}{*}{2015--2016: 6 jets, $\ET > 45~\GeV$}       & ~~~~~~~~~~~2015: 5 jets, $\ET > 70~\GeV$  \\
&                                                           & ~~~~~~~~~~~2016: 5 jets, $\ET > 65~\GeV$  \\
& 2017--2018: 7 jets, $\ET > 45~\GeV$                        & 2017--2018: 5 jets, $\ET > 70~\GeV$ \\\hline
Lepton veto 			        & \multicolumn{2}{c}{0 baseline leptons, $\pt>10~\GeV$} \\\hline
\met significance, \metsig
& \multicolumn{2}{c}{$\metsig>\thrmetsigSR$} \\\hline
\bottomrule
\end{tabular}
\caption{
Summary of common selections prior to the categorisations based on \nbjet and \mjsigma.}
\label{tab:SRdefs}
\end{table}

\begin{table}[h]
\renewcommand\arraystretch{1.4}
\centering
\begin{tabular}{l|c|c|c}
\toprule
 
Signal region & \njfifty  & \nbjet & \mjsigma [\GeV] \\
\midrule
\texttt{SR-8ij50~multi-bin}  & ~~$\geq$8   & \multirow{3}{*}{~~~~~$=0,~ =1,~ \ge2$} & \multirow{3}{*}{~~~~~$(0,340],~ (340,500],~ (500,\infty)$} \\ \cline{1-2}
\texttt{SR-9ij50~multi-bin}  & ~~$\geq$9   & &  \\  \cline{1-2}
\texttt{SR-10ij50~multi-bin} & $\geq$10   & & \\
\bottomrule
\end{tabular}
\caption{Signal region subdivisions for the multi-bin fit in the {\texttt \{8,9,10\}ij50} analysis channels. The selections on these variables apply to the signal regions as well as to the multijet template and validation regions described later in \Sect{\ref{sec:bkgest}}. For each inclusive jet-multiplicity fit, nine regions are generated combining the \nbjet and \mjsigma selections.}
\label{tab:SRVRSummary}
\end{table}

\begin{table}[h]
\renewcommand\arraystretch{1.4}
\centering
\begin{tabular}{l|cccc}
\toprule
Signal region & \njfifty & \njeighty & \nbjet & \mjsigma [\GeV] \\ \midrule
\texttt{SR-8ij50-0ib-MJ500} & ~~$\geq$8 & - & - & $\geq$500 \\
\texttt{SR-9ij50-0ib-MJ340} & ~~$\geq$9 & - & - & $\geq$340 \\
\texttt{SR-10ij50-0ib-MJ340} & $\geq$10 & - & - & $\geq$340 \\
\texttt{SR-10ij50-0ib-MJ500} & $\geq$10 & - & - & $\geq$500 \\
\texttt{SR-10ij50-1ib-MJ500} & $\geq$10 & - & $\geq$1 & $\geq$500 \\
\texttt{SR-11ij50} & $\geq$11 & - & - & - \\
\texttt{SR-12ij50-2ib} & $\geq$12 & - & $\geq$2 & - \\
\texttt{SR-9ij80} & - & $\geq$9 & - & - \\ \hline
\bottomrule
\end{tabular}
\caption{
Summary of signal region criteria for single-bin selections. The requirements in multiplicity and \mjsigma are inclusive in all cases. A dash (`-') indicates that no requirement is applied to the corresponding variable.
The requirement $\metsig>5$ is applied to all bins.}
\label{tab:SRdefsSingleBins}
\end{table}
 
% End of text imported from the .//source/objevtsel.tex input file
 
% The next lines are included from the .//source/bkgest.tex input file
\section{Background estimation}
\label{sec:bkgest}
 
The signal selection in this analysis demands higher jet multiplicities than can practically be generated with fixed-order multi-leg matrix element MC generators for most SM background processes.
The kinematic requirements also challenge the parton shower approximations that are used to obtain higher multiplicities than are possible in fixed-order calculations.
For this reason the analysis was designed in a manner that allows the expected yields in the signal regions for each of the most important backgrounds to be determined either from fully data-driven methods or by using measurements of ATLAS data to calibrate and to confirm the predictions of the MC simulations.
 
Three different broad classes of backgrounds are identified, and a different strategy is employed for each.
 
The first class of background is multijet production through QCD processes in which detector resolution effects produce an apparent momentum imbalance even in the absence of any true \met{}.
This class includes any contribution to the \met{} from neutrinos when semileptonic decays of $b$- and $c$-hadrons are present in those jets.
The probability distribution of jet mismeasurement from either source can be measured with high statistical precision in data events with lower jet multiplicity.
The distribution of \metsig{} is observed to be almost independent of the jet multiplicity, and so the expected yield in each signal region can be determined using a data-driven background template of \metsig{} extracted from lower jet-multiplicity data events.
\Sect{\ref{sec:bkgest:qcdtemplate}} describes this method fully.
 
The second class of backgrounds consists of processes which contain true momentum imbalance caused by the production of neutrinos in association with charged leptons (excluding semileptonic $b$- and $c$-hadron decays, which fall in the first class).
The dominant backgrounds in this second class are from \ttbar{} and \Wjets production.
For such backgrounds, auxiliary measurements of the data in control regions (CRs) are used to normalise the predictions from MC simulation. This control region approach is described in detail in \Sect{\ref{sec:bkgest:leptonic}}.
 
The third class comprises subdominant backgrounds such as events from leptonically decaying $Z$ bosons and associated jets (\Zjets), single-top-quark production, diboson production of $WW$, $WZ$, or $ZZ$, and the production of \ttbar in association with heavier particles -- this final item covers the production  of three or four top quarks, $\ttbar+W$, $\ttbar+Z$, and  $\ttbar+WW$ production, and is collectively denoted by `$\ttbar+X$'.
The expected yields from these backgrounds are found to be sufficiently small that they may be estimated with adequate precision from MC simulation alone.
 
The final determination of the yield from each of the first two classes of backgrounds is obtained using a simultaneous fit including control regions, as described in \Sect{\ref{sec:bkgest:bkgfit}}.
 
\subsection{Multijet background}
\label{sec:bkgest:qcdtemplate}
 
The analysis is designed such that a data-driven method can be used for the estimation of the dominant multijet background.
This technique relies on the approximate invariance of the multijet \metsig distribution with respect to changes in the jet multiplicity.
The procedure is also referred to as the \textit{multijet template estimate} and is schematically shown in \Fig{\ref{fig:bkgest:template_schematic}}, which illustrates various template regions (\texttt{TR}s) used to extract the prediction.
\begin{figure}[h]
\centering
\hspace{-2cm}
\resizebox{0.6\textwidth}{!}{
 
\includegraphics{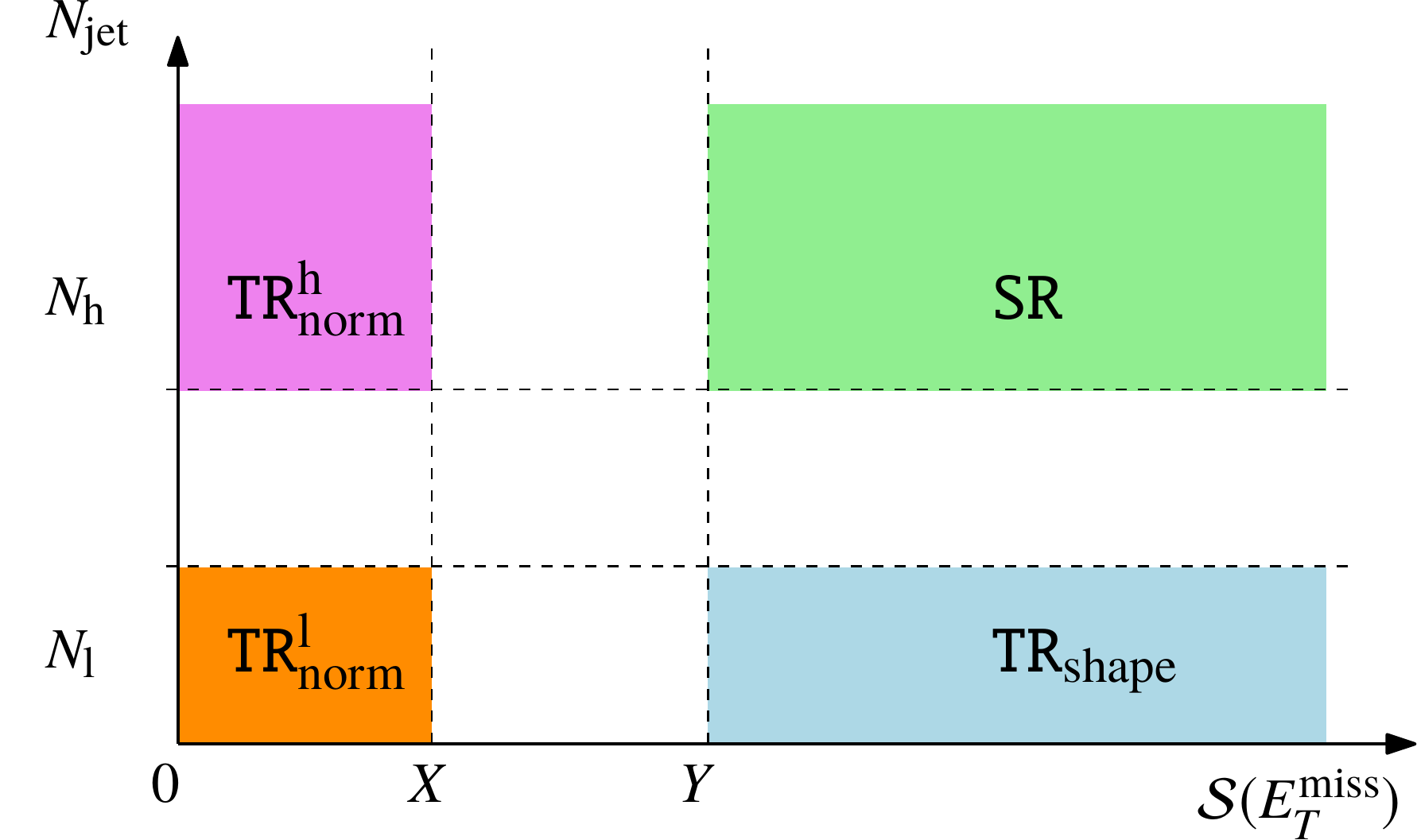}
}
\caption{Schematic of the kinematic regions used in the multijet template estimate for the multijet background. \texttt{SR} represents any signal region, while the three \texttt{TR} variations are regions used to extract and normalise the corresponding background prediction.}
\label{fig:bkgest:template_schematic}
\end{figure}
 
The predicted multijet background yield in a region $\hat{N}[a<\metsig<b]$ with high jet multiplicity ($N_{\textrm{h}}$) and \metsig in the range $(a,b)$ is obtained from the measured yield $N_{\TRshape}$ in a lower jet-multiplicity ($N_{\textrm{l}}$) template region \TRshape through the relation
\begin{equation*}\label{eq:abcd}
\hat{N}[a<\metsig<b] = \frac{N_{\TRnormh}}{N_{\TRnorml}} N_{\TRshape}[a<\metsig<b].
\end{equation*}
In this equation, the regions for template normalisation \TRnorml and \TRnormh, at lower and higher jet multiplicity respectively, require $\metsig < 2$ in order to reduce signal and leptonic background contamination.
The fraction of signal contamination for non-excluded parameter values in the target models is found to range from $<0.1\%$ to $10\%$ depending on the region.
 
The multijet yields in the template regions \TRshape, \TRnorml and \TRnormh are extracted by subtracting the predicted contributions from the two leptonic background classes from the yield in data.
For signal regions requiring $\njfifty\ge 8$--$12$, the \TRnorml and \TRshape are defined using events with $\njfifty=7$.
The criterion $\njeighty=5$ defines the corresponding template regions for the signal region selecting events with $\njeighty\geq 9$.
The \TRnormh selection requires the same jet multiplicity as the corresponding signal region.
 
A residual dependence of the multijet template estimate on event kinematics was observed previously~\cite{SUSY-2015-07}. This secondary dependence is addressed by applying the template method differentially in bins of the total scalar sum of jet transverse momenta $\HT = \sum_{j \in \text{jets}} {\pt}_j$, where ${\pt}_j > \SI{40}{\GeV}$.
The \HT bin ranges are chosen dynamically to ensure good statistical precision in each bin.
The effect of a different choice of the binning on the background estimate is accounted for by a dedicated systematic uncertainty.
For each signal region bin, the template estimate is extracted after applying the same \mjsigma and \nbjet selections to all relevant template regions.
Residual inaccuracies due to the \njet extrapolation and potential mismodelling of the simulated backgrounds in the template regions are corrected in a control region (\QCR) defined in the range $3<\metsig<4$, and at the same jet multiplicity as the signal region.
The yield in each \QCR is used as input to the simultaneous fit described in \Sect{\ref{sec:bkgest:bkgfit}}.
 
To verify the accuracy of the multijet background prediction, two sets of validation regions are defined.
These correspond to the two dimensions in which the background prediction is extrapolated:
\begin{itemize}
\item \VRnjet is defined at a jet multiplicity below that of the signal region, and tests the extrapolation in \metsig;
\item \VRmetsig instead is defined in an intermediate \metsig range between the \QCR and the signal region, specifically $4<\metsig<5$, and tests the extrapolation in \njet.
\end{itemize}
 
For the \texttt{SR-9ij80} signal region, the \VRnjet requires $\njeighty=6$, a multiplicity between the template regions and the signal region selection.
An intermediate jet-multiplicity region is not available for the \texttt{j50} selections, as the \texttt{SR-8ij50} regions are adjacent to the template regions.
Instead, the \VRnjet requires  $\njfifty=7$, and the test is performed on a template extracted from data with $\njfifty=6$, selected with a prescaled trigger.
The template region \metsig and \njet requirements are summarised together with the \QCR and \VR requirements in \Tab{\ref{tab:bkgest:TRdefsj80}} and \Tab{\ref{tab:bkgest:TRdefsj50}}. Example distributions of \metsig in the lower jet-multiplicity {\VR}s are shown in Figure~\ref{fig:bkgest:validation}.
 
\begin{table}[htbp]
\renewcommand\arraystretch{1.2}
\centering
\begin{tabular}{l|ccccc}
\toprule
\metsig range & 0--2 & 2--3 & 3--4 & 4--5 & $>5$ \\
\midrule
$\njeighty=5$ & \TRnorml & -- & -- & -- & \TRshape \\
$\njeighty=6$ & -- & -- & -- & -- & \VRnjet \\
$\njeighty\ge9$ & \TRnormh & -- & \QCR & \VRmetsig & \SR \\
\bottomrule
\end{tabular}
\caption{
Illustration of the main multijet template, control and validation regions in \njet and \metsig corresponding to the \texttt{SR-9ij80} signal region.
The template regions used to derive predictions in the control and validation regions are not shown.
}
\label{tab:bkgest:TRdefsj80}
\end{table}
 
\begin{table}[htbp]
\renewcommand\arraystretch{1.2}
\centering
\begin{tabular}{l|ccccc}
\toprule
\metsig range & 0--2 & 2--3 & 3--4 & 4--5 & $>5$ \\
\midrule
$\njfifty=6$, prescaled data & $\mathtt{TR}_\mathrm{norm}^{\mathrm{l,}\mathtt{prescale}}$ & -- & -- & -- & $\mathtt{TR}_\mathrm{shape}^\mathtt{prescale}$ \\
$\njfifty=7$, full dataset & $\mathtt{TR}_\mathrm{norm}^{\mathrm{h,}\mathtt{prescale}}$  & -- & -- & -- & \VRnjet \\
\midrule
$\njfifty=7$, full dataset & \TRnorml & -- & -- & -- & \TRshape \\
$\njfifty\ge8$, full dataset & \TRnormh & -- & \QCR & \VRmetsig & \SR \\
\bottomrule
\end{tabular}
\caption{
Illustration of the main multijet template, control and validation regions in \njet and \metsig corresponding to the \texttt{j50} signal regions.
The template regions needed to derive predictions in the control and \VRmetsig regions are not shown.
For the row labelled `prescaled data', a six-jet trigger was used that collected only a fraction of the Run~2 data.
The superscript `\texttt{prescale}' is used to indicate the template regions used to predict the \VRnjet background using prescaled data.
}
\label{tab:bkgest:TRdefsj50}
\end{table}
 
\begin{figure}[htbp]
\centering
\subfloat[$N_\mathrm{jet}^{50} = 7$, $N_{b\mathrm{-jet}}\geq2$]{\includegraphics[width=0.49\textwidth]{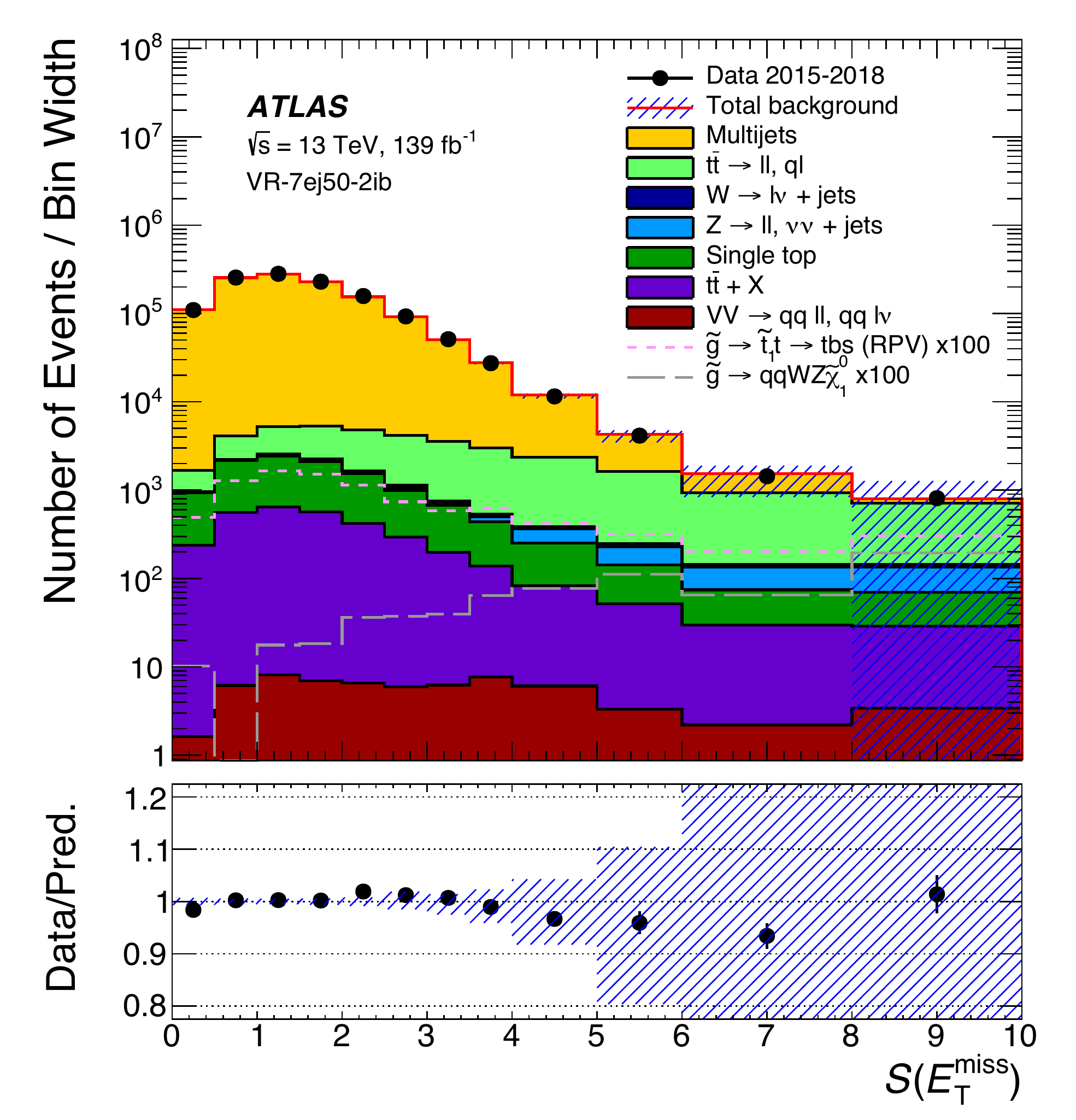}}
\hspace{0.5ex}
\subfloat[$N_\mathrm{jet}^{80} = 6$, $N_{b\mathrm{-jet}}\geq0$ ]{\includegraphics[width=0.49\textwidth]{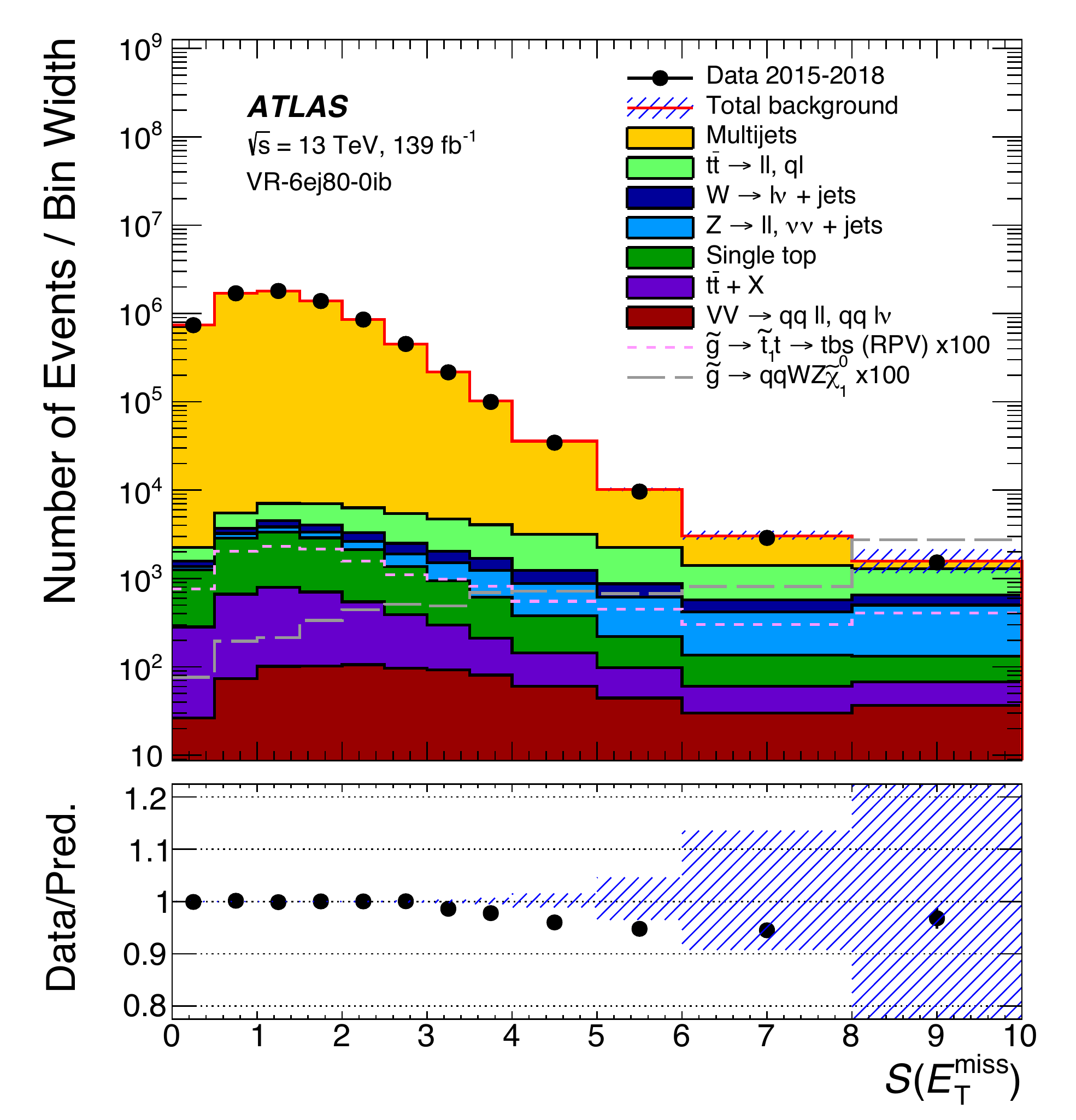}}
\caption{Distributions of \metsig for events in two of the validation regions.
\dataMCcaption\,\signalcaption\,
Signal yields are scaled by a factor of 100 for visibility.
\ratiopanelcaption\,
}
\label{fig:bkgest:validation}
\end{figure}
 
\FloatBarrier
 
\subsection{Leptonic backgrounds}
\label{sec:bkgest:leptonic}
 
The SM processes which produce multijet events with one or more leptons are categorised as \textit{leptonic} backgrounds.
Although events containing charged leptons ($e$ or $\mu$) are excluded from the signal regions, it is still possible for leptonic backgrounds to contaminate them.
The veto only applies to events containing electrons or muons, and hence hadronically decaying $\tau$-leptons (originating from top quark or $W$ boson decays) remain a source of background.
Such $\tau$-leptons are treated as jets within this analysis, so they may contribute to the jet count if they have sufficient \pT{}, and the momentum lost through any associated neutrinos can also cause these events to enter the signal regions.
Additionally, there are cases where the electrons or muons may fall outside of the detector acceptance, or be misreconstructed in the detector, increasing the $\met$ of the event.
 
The two largest leptonic backgrounds are from leptonically and semileptonically decaying \ttbar, and leptonically decaying $W$ bosons produced in association with jets. The estimation of these backgrounds employs the MC simulations described in Section~\ref{sec:datasets}. To reduce normalisation and modelling uncertainties the background predictions are normalised to data using CRs.
 
The control regions are designed to be kinematically similar to signal regions, but not to overlap with them. They are designed to enhance the contributions from  particular backgrounds, in order to measure those backgrounds cleanly, while being comparatively sparse in signal contamination.
Statistical orthogonality between the signal regions and the leptonic CRs is achieved by requiring exactly one electron or muon in CR events.
To reduce statistical uncertainties, each $\Njet\geq m$ signal region has a corresponding ($\Njet\geq m-1$) control region, except in the case of signal regions with an $\njfifty\ge8$ requirement, where there are sufficiently large yields in the control region to match the signal region $\Njet\geq m$ requirement.
In order to increase the statistical precision in control regions with higher jet-multiplicity requirements, the \metsig{} threshold is reduced to $4$ in the leptonic control regions.
 
Two leptonic control regions are defined for each signal region: the first (\WCR) includes a \bjet veto to enhance the contribution of \Wjets backgrounds, and the second (\TCR) requires at least one \bjet in the event, enriching the region with the \ttbar background. The signal-region \pt thresholds imposed on the jets also apply to the corresponding control regions.
To simulate the effect of $\tau$-leptons being reconstructed as jets in the signal regions, electrons and muons are also included as jets, for the purpose of the corresponding CR selection, provided that they pass the same \pt and $|\eta|$ requirements as the jets in the event.
Finally, an upper bound on the transverse mass\footnote{The transverse mass is defined as $ \mt  =  \sqrt{2\pt^\ell  \met \left[ 1-\cos \left( \Delta\phi(\vec{\pt}^\ell,\vec{E}{_\text{T}^{\text{miss}}})  \right) \right]} $, where $\pt^\ell$ is the lepton $\pt$.} computed with the lepton and \met is applied at 120~\GeV\ in order to reduce contributions from signal processes. This variable has a kinematic endpoint at the $W$ boson mass for leptonically decaying on-shell $W$ bosons, but has no such bound when the lepton and $\met$ originate from different decays.
The control region definitions are summarised in \Tab{\ref{tab:bkgest:ControlRegions}}.
 
Additional requirements are placed on \mjsigma for the control regions in the same manner as for the corresponding signal region selections.
In the case of the single-bin selections (\Tab{\ref{tab:SRdefsSingleBins}}), the same threshold is applied.
For multi-bin selections, the control regions have three bins, corresponding to the same \mjsigma thresholds, as shown in \Tab{\ref{tab:bkgest:WTCRSummary}}.
 
The \WCR and \TCR are used as inputs to a fit that applies normalisation corrections to the \ttbar and \Wjets background components, as described in \Sect{\ref{sec:bkgest:bkgfit}}.
 
\begin{table}[h]
\renewcommand\arraystretch{1.2}
\centering
\begin{tabular}{l|c}
\toprule
Selection criterion & Selection ranges \\
\midrule
Lepton multiplicity & Exactly one signal $e$ or $\mu$ remaining after overlap removal \\
Lepton $\pt$ & $>20~\GeV$ \\
\mt & $<120~\GeV$ \\
Trigger & Same as signal regions \\
Jet \pt, $|\eta|$ & Same as signal regions \\
\multirow{2}{*}{\Njet (including lepton)} & $\ge8$, for $\njfifty\ge8$ signal regions; \\
& $\ge (N_\mathrm{jet}^\SR - 1)$, otherwise~~~~~~~~~~~~~~ \\
$\Nbjet$ & = 0 (\WCR), $\geq$ 1 (\TCR) \\
\mjsigma & Same as signal regions \\
$\metsig$ & $>~4$ \\
\bottomrule
\end{tabular}
\caption[Control regions]{Summary of the selections used to define the leptonic control regions.}
\label{tab:bkgest:ControlRegions}
\end{table}
 
\begin{table}[h]
\renewcommand\arraystretch{1.2}
\centering
\begin{tabular}{c|cc}
\toprule
8, 9, 10ij50 & \multirow{2}{*}{\nbjet = 0} & \multirow{2}{*}{\nbjet $\ge1$}\\
CR definitions &&  \\
\midrule
$\mjsigma\leq 340~\GeV$ & WCR1 & TCR1 \\
$340~\GeV < \mjsigma \leq 500~\GeV$& WCR2 & TCR2 \\
$\mjsigma > 500~\GeV$ & WCR3 & TCR3\\
\bottomrule
\end{tabular}
\caption{Control region subdivisions for the background fits for multi-bin signal selections.}
\label{tab:bkgest:WTCRSummary}
\end{table}
 
\subsection{Background normalisation corrections}
\label{sec:bkgest:bkgfit}
 
Background estimates in the signal region are made more accurate by employing a background likelihood fit based on the control regions, using the methods described in Ref.~\cite{Baak:2014wma}.
By means of this fit, the raw estimated yields for the major background components,
including the multijet, \ttbar, and \Wjets processes, are corrected for mismodelling.
For the other background processes, which contribute of the order of 1\% of the SR yields, the nominal MC predictions are used directly.
The extent to which the background prediction is compatible with the signal region observation is quantified in the form of a $p$-value $\mathrm{CL_b}$, which is the probability of an upward fluctuation of the event yield relative to the signal region prediction no larger than that observed in data, given the background model.
The multijet background, while estimated using a data-driven procedure, is incorporated in the simultaneous fit due to the dependence of the template prediction on the subtraction of other backgrounds which include the simulated \ttbar and \Wjets estimates.
Besides correcting for any residual mismodelling, fitting the multijet component handles the correlations between the systematic uncertainties of the different background components consistently.
 
In the case of the single-bin regions (\Tab{\ref{tab:SRdefsSingleBins}}), the normalisations of the background components are allowed to vary within their nominal uncertainties, described in \Sect{\ref{sec:systematics}}.
For the multi-bin analysis channels (\Tab{\ref{tab:SRVRSummary}}), the additional information available to the fit permits a reduction in the uncertainties, as well as a modification of the event yields to better accommodate the control region measurements.
To avoid artificially constraining systematic uncertainties in the background fit due to the high statistical precision, the multijet normalisation region \QCR is limited to a single bin with $\nbjet\ge0$ and $\mjsigma\ge0$.
 
For illustration, the pre-fit yields for the \texttt{SR-8ij50} leptonic CRs are shown in \Fig{\ref{fig:bkgest:fitregions_8ij50_multibin}}, demonstrating the extent of the observed mismodelling in the \Wjets and \ttbar normalisation.
The fitted normalisation factors are summarised for all signal selections in \Fig{\ref{fig:bkgest:mu_summary}}, and are found to be consistent across the wide range of jet multiplicities probed.
 
\begin{figure}
\centering
\subfloat[WCR, pre-fit]{\includegraphics[width=0.49\textwidth]{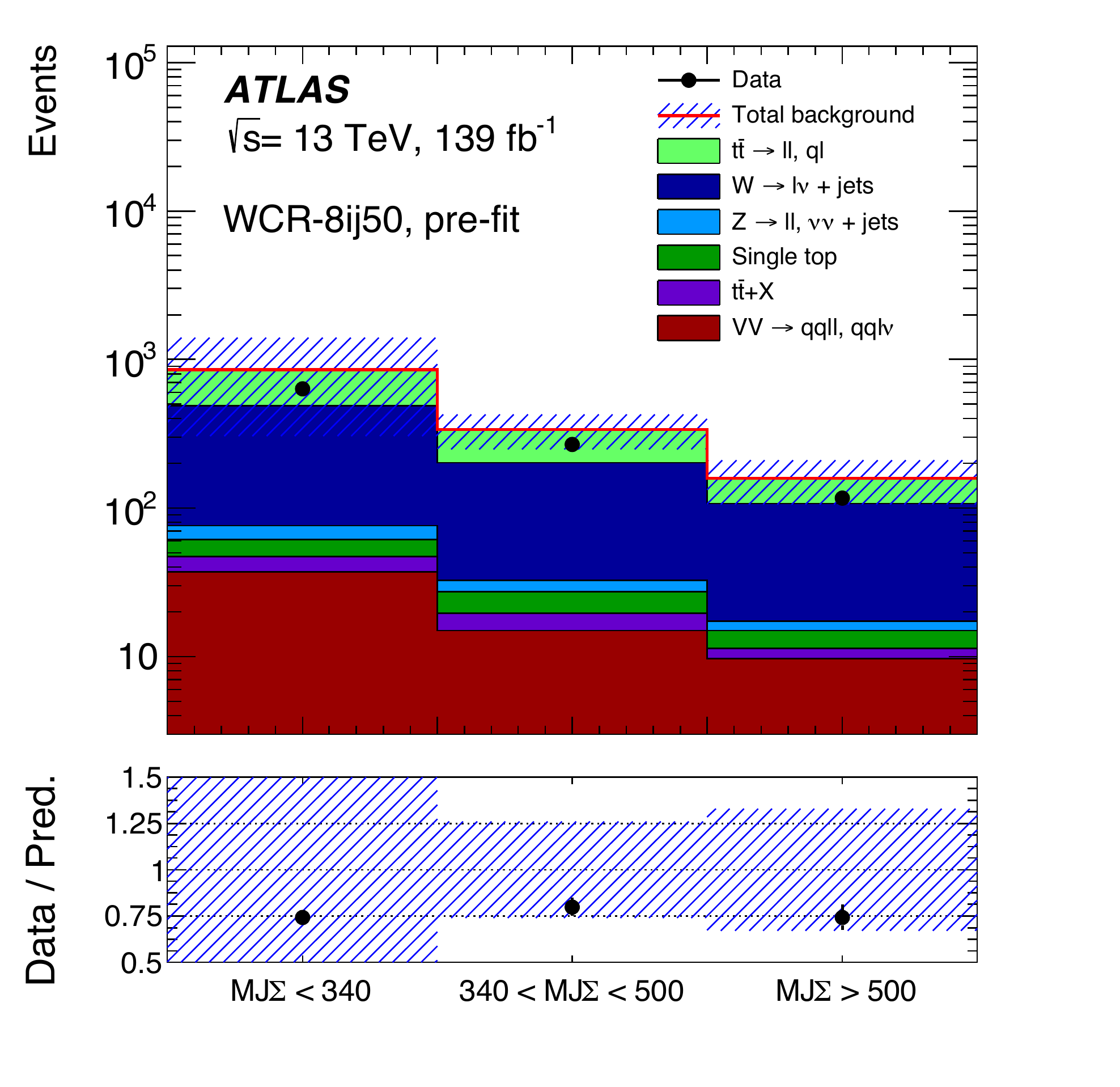}}
\hspace{0.5ex}
\subfloat[TCR, pre-fit]{\includegraphics[width=0.49\textwidth]{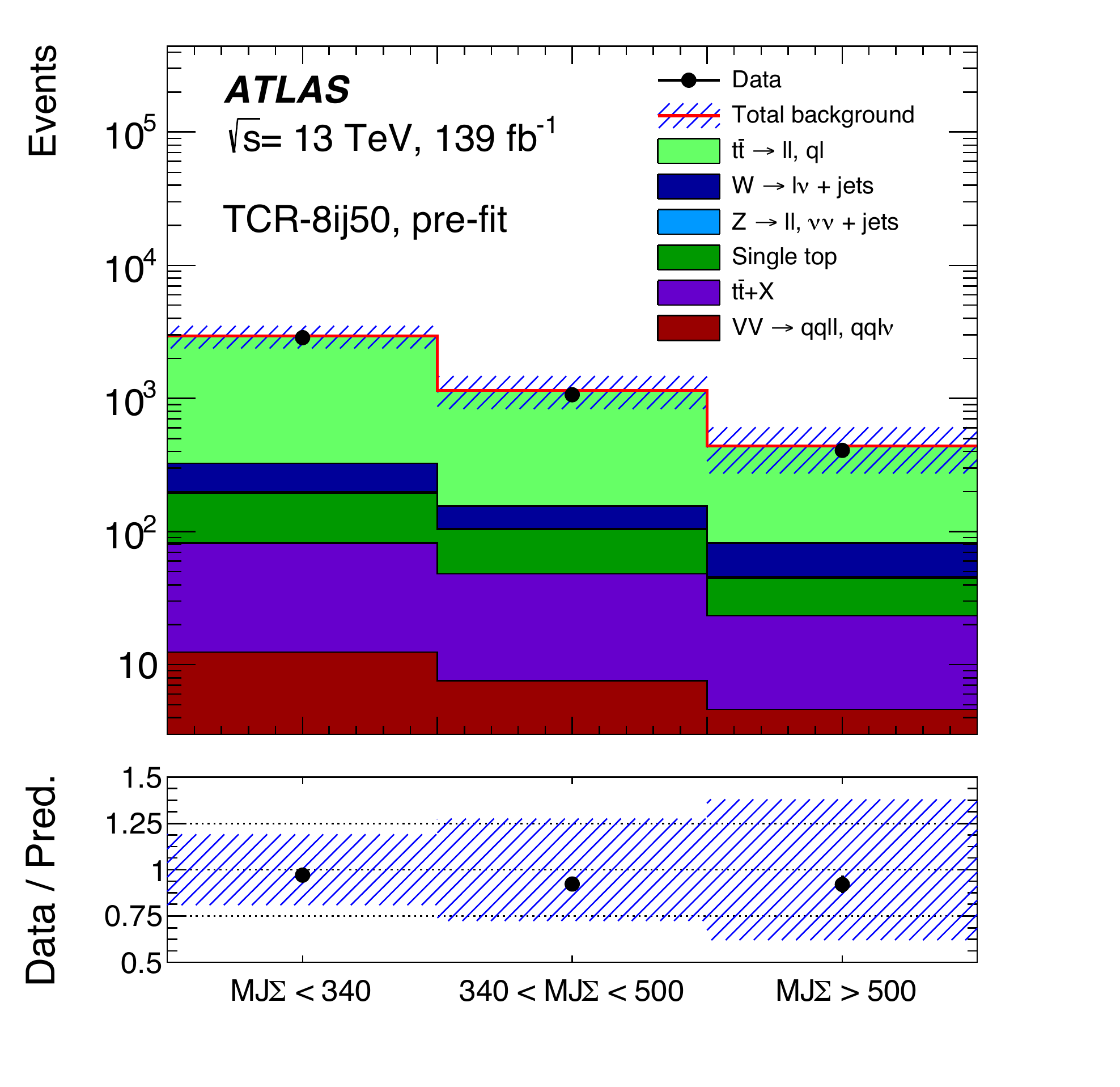}}
\caption{
Pre-fit yields in the (a) \Wjets\ and (b) \ttbar\ background normalisation regions for the \texttt{SR-8ij50} multi-bin analysis.
\dataMCcaption\,\ratiopanelcaption\,}
\label{fig:bkgest:fitregions_8ij50_multibin}
\end{figure}
 
\begin{figure}
\centering
\includegraphics[width=0.9\textwidth]{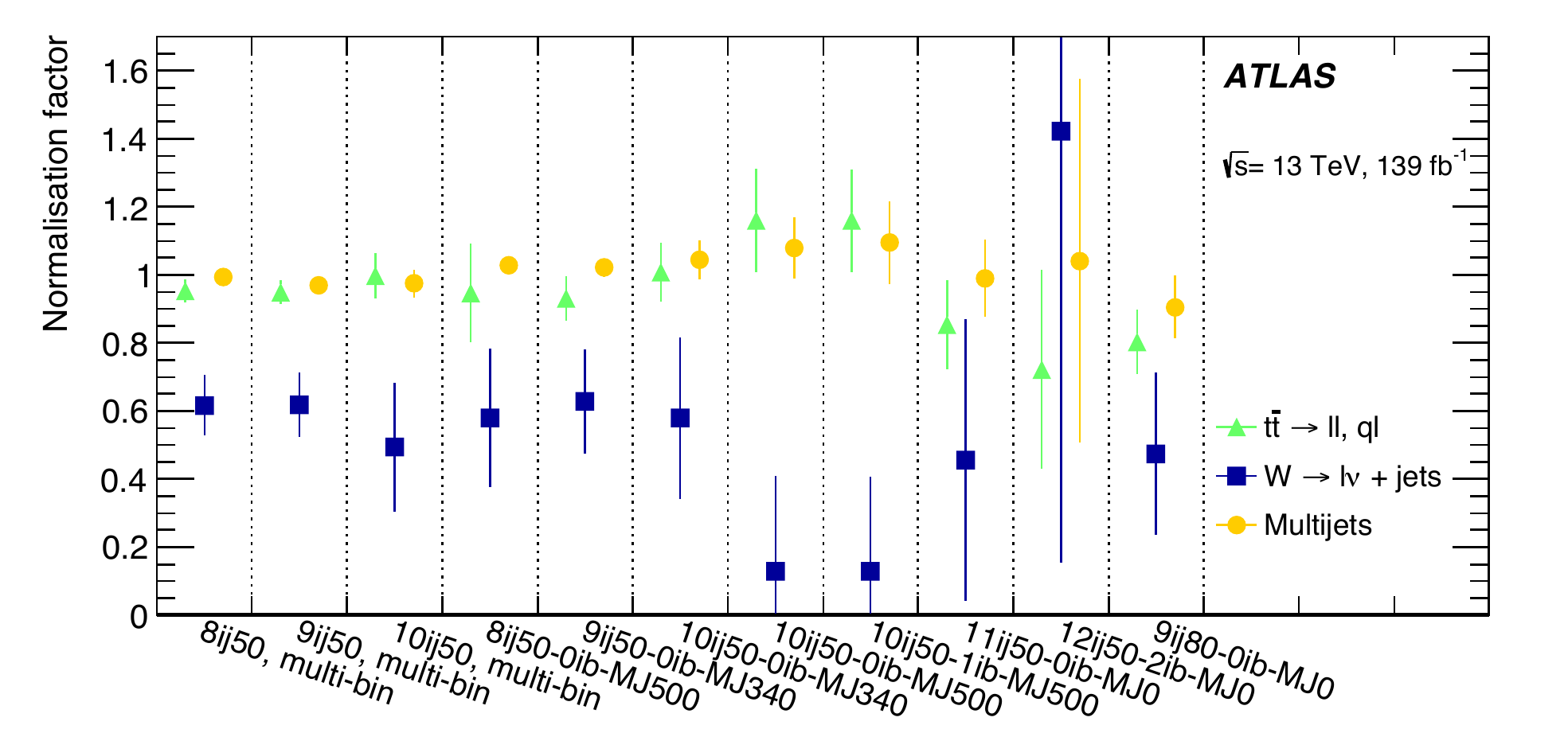}
\caption{Summary of the fitted normalisation factors for the \ttbar, \Wjets and multijet backgrounds in all signal regions.
The error bars indicate the combination of statistical and systematic uncertainties in the corresponding factors.
Two pairs of SRs, namely the \texttt{SR-8ij50} and \texttt{SR-9ij50} multi-bin regions and the \texttt{SR-10ij50-0ib-MJ500} and \texttt{SR-10ij50-1ib-MJ500} single-bin regions share leptonic control regions, and therefore have highly correlated normalisation factors for the \Wjets\ and \ttbar\ background components.}
\label{fig:bkgest:mu_summary}
\end{figure}

\Fig{\ref{fig:bkgest:VRs_8ij50_multibin}} shows the background modelling in the validation regions \VRnjet and \VRmetsig for the \texttt{SR-8ij50} multi-bin analysis.
Considering uncertainties, the data yields are in agreement with the predictions after applying the background normalisation factors.
Similar levels of agreement are found for the single-bin signal regions, as can be seen in \Fig{\ref{fig:bkgest:VRs_singlebin}}.
In the \VRmetsig, there is a tendency for the background predictions to mildly overshoot the data, at the level of 10\%.
This is due to residual kinematic correlations causing the \metsig distribution not to be entirely independent of the jet multiplicity.
Applying flavour-tagging and jet mass selections alters these correlations.
In the validation and signal regions, an uncertainty based on the largest observed non-closure at lower jet multiplicities or smaller \metsig\ values is applied, and is found to cover the observed discrepancies.
 
\begin{figure}
\centering
\subfloat[$\texttt{VR}_{N_\mathrm{jet}}$]{\includegraphics[width=0.49\textwidth]{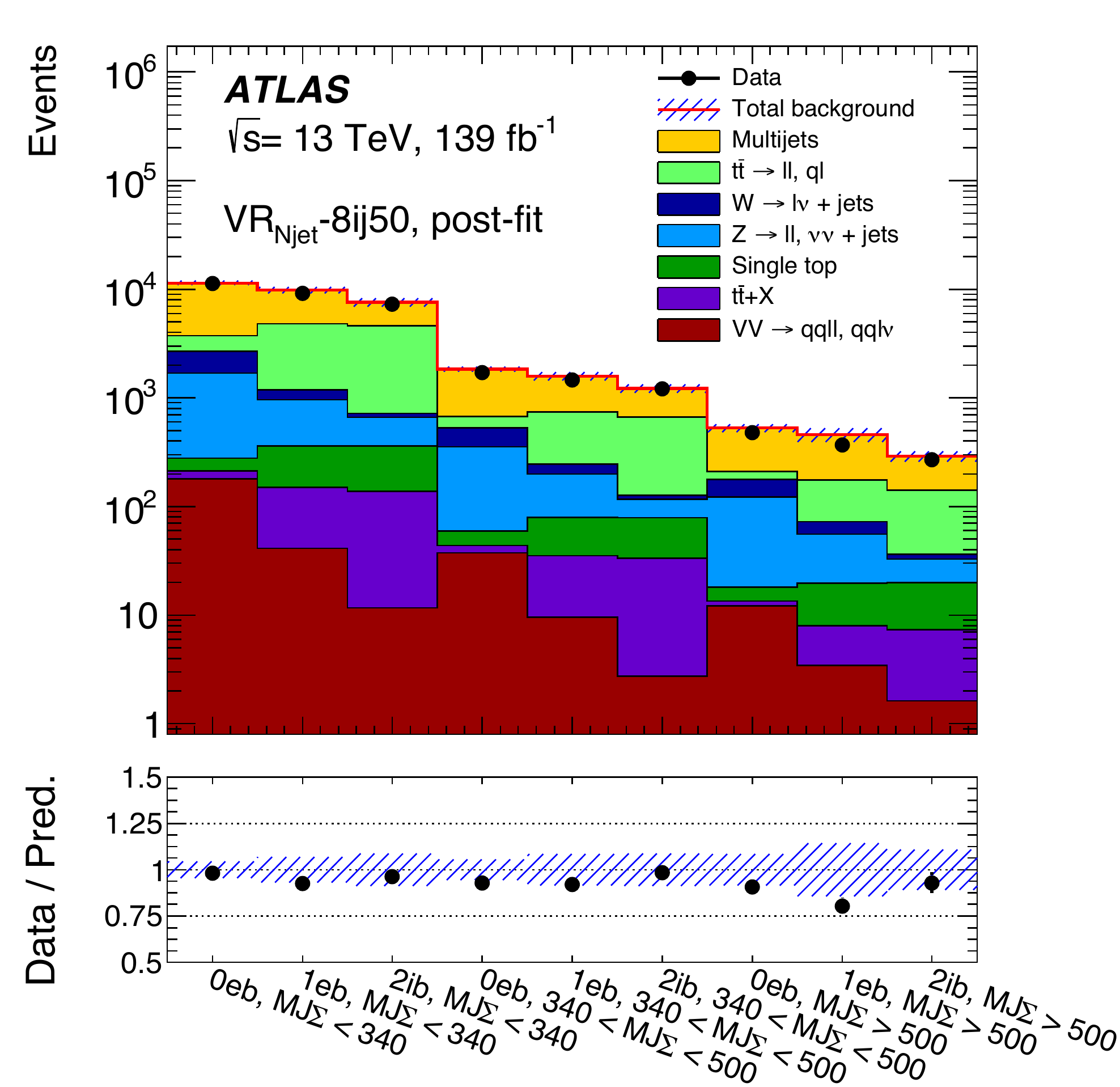}}
\subfloat[$\texttt{VR}_{\mathcal{S}{(E_{T}^\mathrm{miss})}}$]{\includegraphics[width=0.49\textwidth]{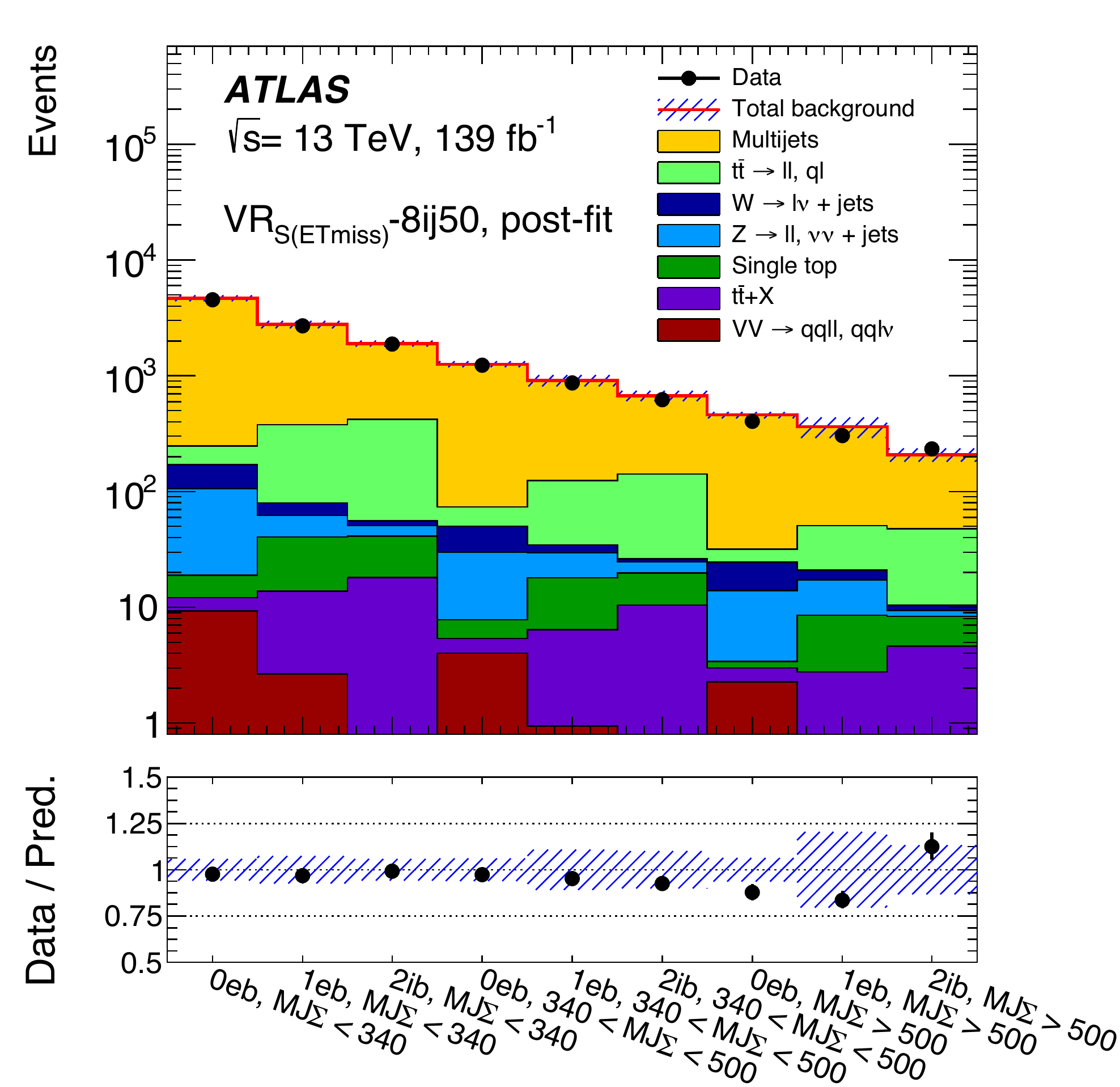}}
\caption{
Post-fit event yields in validation regions for the \texttt{SR-8ij50} multi-bin analysis.
\dataMCcaption\,\ratiopanelcaption\,}
\label{fig:bkgest:VRs_8ij50_multibin}
\end{figure}
 
\begin{figure}
\centering
\subfloat[$\texttt{VR}_{N_\mathrm{jet}}$]{\includegraphics[width=0.49\textwidth]{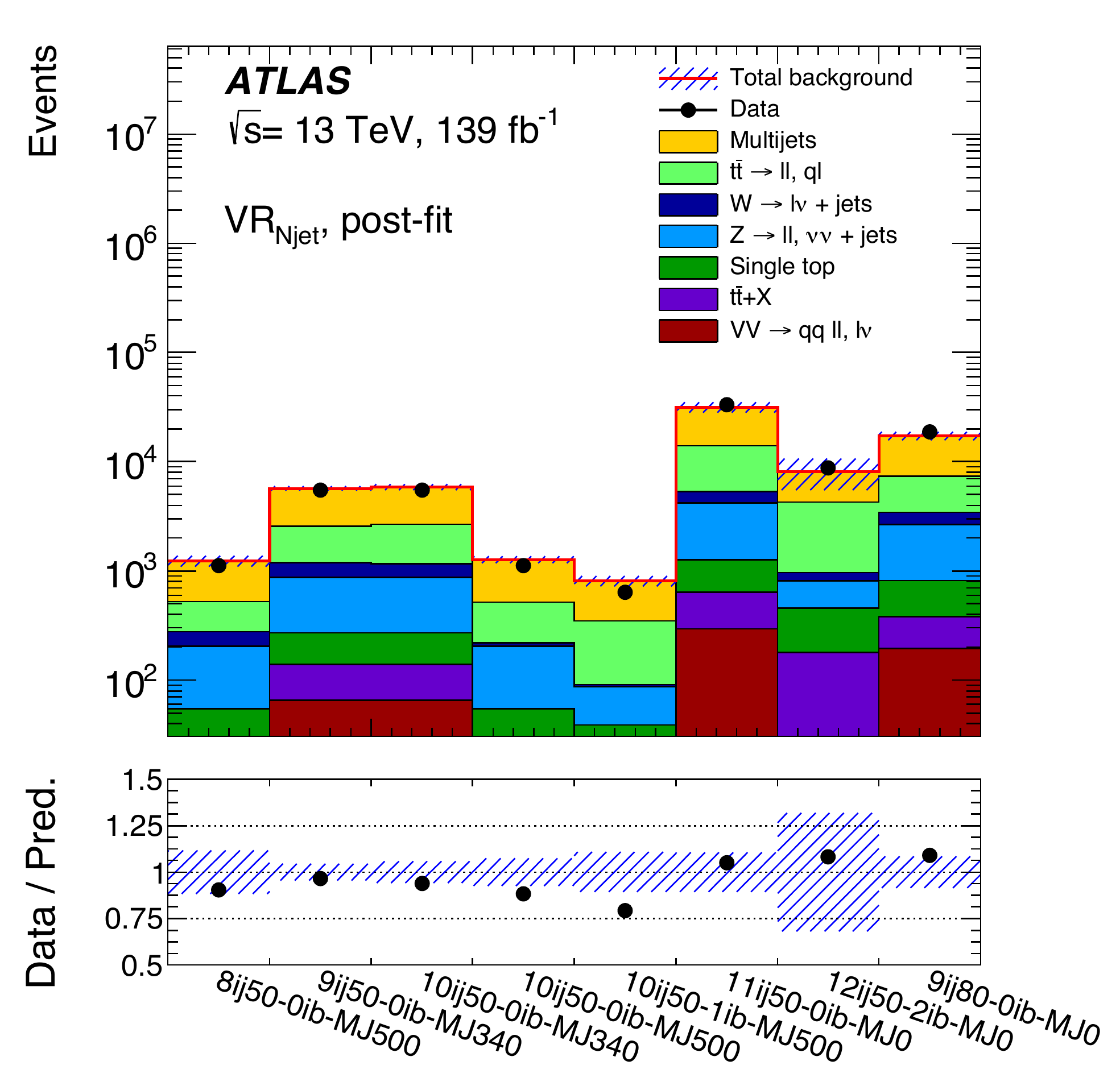}}
\hspace{0.5ex}
\subfloat[$\texttt{VR}_{\mathcal{S}{(E_{T}^\mathrm{miss})}}$]{\includegraphics[width=0.49\textwidth]{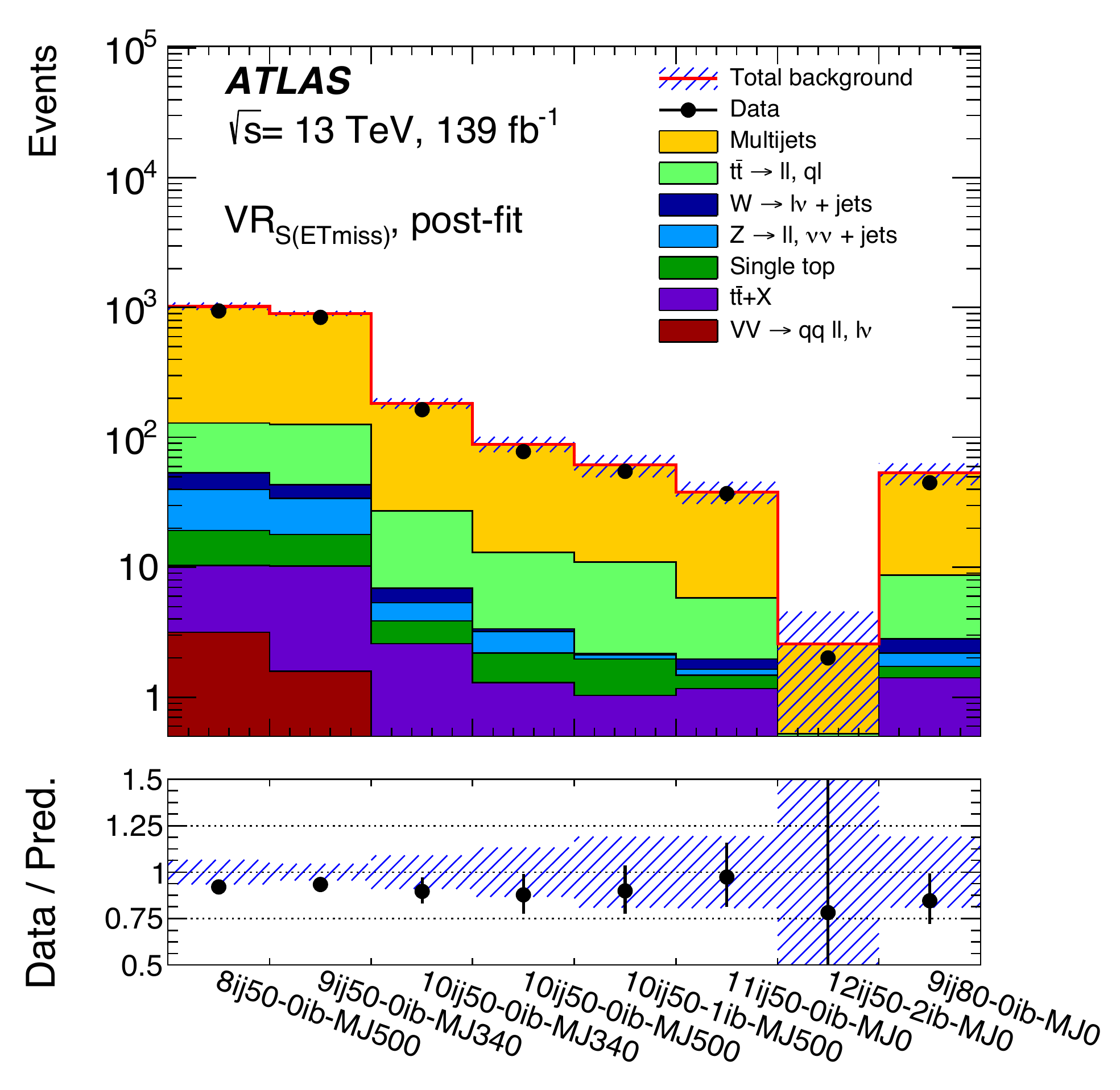}}
\caption{
Post-fit event yields in validation regions summarised for all single-bin signal regions.
\dataMCcaption\,\ratiopanelcaption\,}
\label{fig:bkgest:VRs_singlebin}
\end{figure}

\subsection{Systematic uncertainties}
\label{sec:systematics}
 
Systematic uncertainties from the following sources are assessed for the predicted background yields.
Experimental systematic uncertainties chiefly include uncertainties in the energy or momentum scales of reconstructed jets and leptons or the missing transverse momentum, as well as the uncertainty in the total integrated luminosity and the magnitude of pile-up corrections.
Theoretical uncertainties are assessed by varying the scales (renormalisation, factorisation, resummation, shower matching) at which cross-sections are calculated or by comparison of an ensemble of matrix element and parton shower programs used to generate the predictions.
Each source of theoretical uncertainty is correlated across the signal, control and validation region selections, but assessed separately for each background process.
 
Additional uncertainties account for potential inaccuracies in the data-driven multijet estimate. The effects of residual kinematic correlations are estimated by modifying the \HT-binning procedure used in the multijet estimate. A comparison between the nominal prediction and an alternative prediction assuming a broader resolution for flavour-tagged jets measured in data is used to estimate the impact of different flavour composition in the template and signal regions. Finally, a conservative uncertainty is assessed from the observed non-closure of the prediction in validation regions.
 
The dominant sources of systematic uncertainty are listed in \Tab{\ref{tab:leadingsyst}}.
The total uncertainty in the predicted signal region yield is in the range
6--14\% for the multi-bin signal regions, and 7--66\% for the
less constrained single-bin regions.
In both cases the largest
uncertainties arise in the regions with the highest requirements on the jet multiplicity
(and in the single-bin case also on the $b$-jet multiplicity or \mjsigma{})
since in those cases the supporting measurements are more statistically limited
and so provide less precise constraints on the background predictions.
 
In each of the three multi-bin signal regions the most important uncertainty is the closure systematic uncertainty associated with the multijet template estimate (3--6\%),
chiefly because of the dominance of this background component.
In the single-bin signal regions the background normalisations
are less constrained by the fit, and the statistical uncertainties associated with the
multijet and $\ttbar$ control regions tend to dominate (4--27\% for multijets, 6--14\% for $\ttbar$).
 
The theoretical uncertainty in the \ttbar background predictions is also found to be significant,
with that from final-state radiation contributing up to 17\% in the single-bin fits.
The impact on the total yield in multi-bin signal regions is $\leq$5\%.
The \Zjets\ component has a theoretical uncertainty of up to 14\% in the single-bin regions while it is at most 6\% in the multi-bin signal regions.
 
Most experimental systematic uncertainties affecting the reconstructed objects have insignificant impacts, being substantially reduced due to correlations between the MC-based predictions and the data-driven multijet template.
The largest observed effects are due to the uncertainties in the jet energy scale, which can have an effect of up to 3\% due to the large jet activity in the events selected by this analysis, and in the soft term of the missing transverse momentum ($\leq$10\%).
In the least populated SRs, there can be a large statistical uncertainty in the predictions from simulation.
 
\begin{table}[h]
\centering
\small
\renewcommand\arraystretch{1.2}
 
\begin{tabular}{l|c|l r|l r|l r}
\hline
Signal region & Total syst. & \multicolumn{6}{c}{Dominant systematic uncertainties} \\
\hline
\texttt{SR-8ij50-0ib-MJ500}     &         16\% & \SYalphaMETSoftTrkResoPara     &   7\% & \SYalphaMETSoftTrkResoPerp            &   7\% & \SYalphaqsfZjets                      &   5\% \\
\texttt{SR-9ij50-0ib-MJ340}     &         16\% & \SYalphaMETSoftTrkResoPerp     &   9\% & \SYalphaMETSoftTrkResoPara            &   9\% & \SYalphaqsfZjets                      &   4\% \\
\texttt{SR-10ij50-0ib-MJ340}&     20\% & \SYalphaFSRttbar                       &   9\% & \SYgammaSR                                            &   9\% & \SYalphaMETSoftTrkResoPara    &   8\% \\
\texttt{SR-10ij50-0ib-MJ500}&     27\% & \SYalphaFSRttbar                       &  17\% & \SYgammaSR                                            &   12\% & \SYalphaMETSoftTrkResoPara   &   9\% \\
\texttt{SR-10ij50-1ib-MJ500}&     24\% & \SYgammaSR                                     &  14\% & \SYalphaMETSoftTrkResoPara    &  10\% & \SYalphaMETSoftTrkResoPerp            &  10\% \\
\texttt{SR-11ij50}              &         27\% & \SYgammaSR                                     &  18\% & \SYalphaFSRttbar                      &  14\% & \SYmuttbar                                            &   6\% \\
\texttt{SR-12ij50-2ib}          &         70\% & \SYgammaSR                                     &  62\% & \SYmumultijet                         &  25\% & \SYalphaHTBinning                             &  13\% \\
\texttt{SR-9ij80}                       &         21\% & \SYgammaSR                                     &  14\% & \SYalphaqsfZjets                      &  14\% & \SYalphackkwZjets                             &   7\% \\
\hline
\texttt{SR-8ij50 multi-bin}     &          ~~6\% & \SYalphanonClosure                   &   3\% & \SYalphaJETFlavourComp                        &   3\% & \SYalphaJETFlavourResp                &   2\% \\
\texttt{SR-9ij50 multi-bin}     &          ~~7\% & \SYalphanonClosure                   &   4\% & \SYalphaqsfZjets                              &   3\% & \SYalphaMETSoftTrkResoPara    &   3\% \\
\texttt{SR-10ij50 multi-bin}&     14\% & \SYalphaqsfZjets                       &   6\% & \SYalphanonClosure                            &   6\% & \SYalphaFSRttbar                      &   5\% \\
\hline
\end{tabular}
\caption{
The total systematic uncertainties are shown for each of the single-bin signal regions,
and also for the multi-bin signal regions,
together with the three dominant contributions for each.
The individual uncertainties can be (anti-)correlated,
and do not necessarily sum in quadrature to the total
background uncertainty.
For the multi-bin signal regions the uncertainties are those found after summing
the expected yields over the corresponding \mjsigma\ and $b$-jet multiplicity bins of \Tab{\ref{tab:SRVRSummary}}.
Within the table
`\SYgammaSR' indicates the statistical uncertainty of the simulated event yield in the SR,
`MJ' indicates the uncertainty in the multijet background,
`closure' indicates the uncertainty from the multijet template method closure,
`norm' is the result of statistical uncertainties from the CRs,
`JES\,flavour' indicates the effect of uncertainties in the jet energy scale due to differences between quark- and gluon-initiated jets,
`\met\ soft, L/T' indicate two sources, longitudinal and transverse, of uncertainty in the soft component of the missing transverse momentum,
`\SYalphaHTBinning' relates to the parameters of the binning of the multijet template in \HT,
`FSR' indicates final-state radiation,
`match' indicates the matrix element/parton shower matching scale,
and `PS' is the uncertainty from varying the scale at which the strong coupling constant
is calculated for parton shower emissions in the MC simulation.
}
\label{tab:leadingsyst}
\end{table}
% End of text imported from the .//source/bkgest.tex input file

% The next lines are included from the .//source/results.tex input file
\section{Results and interpretation}
\label{sec:result}
 
Data yields are shown graphically for all signal regions in \Fig{\ref{fig:results:SRyields}}.
The background predicted to have the largest yield in all signal regions comes from multijet production.
The relative contribution of the remaining backgrounds depends on the signal region.
The \ttbar{} process generally provides the second-largest contribution,
and tends to form a larger fraction of the total background for higher jet-multiplicity requirements.
As the requirement on the number of $b$-tagged jets increases, it can be seen that the
relative contributions of \Wjets{}, \Zjets{} and multi-boson backgrounds decrease compared to
those from \ttbar{} and single-top-quark production processes.
 
A breakdown of the yields in the single-bin regions is given in \Tab{\ref{tab:results:singlebinyields}}.
For illustration, the full \metsig{} distributions for several signal regions are shown in \Fig{\ref{fig:results:SRmetsig}}.
The data yields are found to be consistent with the background predictions within the assessed statistical and systematic uncertainties, with no significant excesses over the SM expectation.
Mild deviations from the SM expectation are observed, with a tendency for the background to be overestimated in the higher jet-multiplicity regions, which is consistent with the trends observed in the corresponding validation regions.
For interpretation, the likelihood fits for background estimation (\Sect{\ref{sec:bkgest:bkgfit}}) are extended to include the signal region, and thereby perform two forms of hypothesis test using a profile-likelihood-ratio test statistic~\cite{Cowan:2010js}, quantifying the significance of any observed excesses or the lack thereof.
 
Firstly, the discovery test discriminates between the null hypothesis stating that the SR measurement is consistent with only SM contributions and an alternative hypothesis postulating a positive signal.
Secondly, assuming a specific signal model, one may also form an exclusion test of the signal-plus-background hypothesis, where an observation significantly smaller than the combination of SM and SUSY processes would lead to rejection of the signal model.
This provides the exclusion $p$-value $p_1$, the probability of observing at most the observed event yield when assuming that the signal is present with its nominal cross-section.
A complementary $p$-value for the background observation $\mathrm{CL}_\mathrm{b}$ is defined as the probability of observing at most the observed yield under the background-only hypothesis.
Points in the SUSY parameter space are considered excluded if the ${\mathrm{CL}}_{\mathrm{s}}$ parameter~\cite{Read:2002hq}, computed as $p_1/(1-{\mathrm{CL}}_{\mathrm{b}})$ is smaller than 0.05.
This protects against spurious exclusion of signals due to observing SR event counts significantly smaller than those predicted. While not strictly
defining a frequentist confidence level, these are referred to as 95\% confidence level (CL) limits.

\begin{figure}[hbp]
\centering
\subfloat[SR-8ij50, multi-bin]{\includegraphics[width=0.49\textwidth]{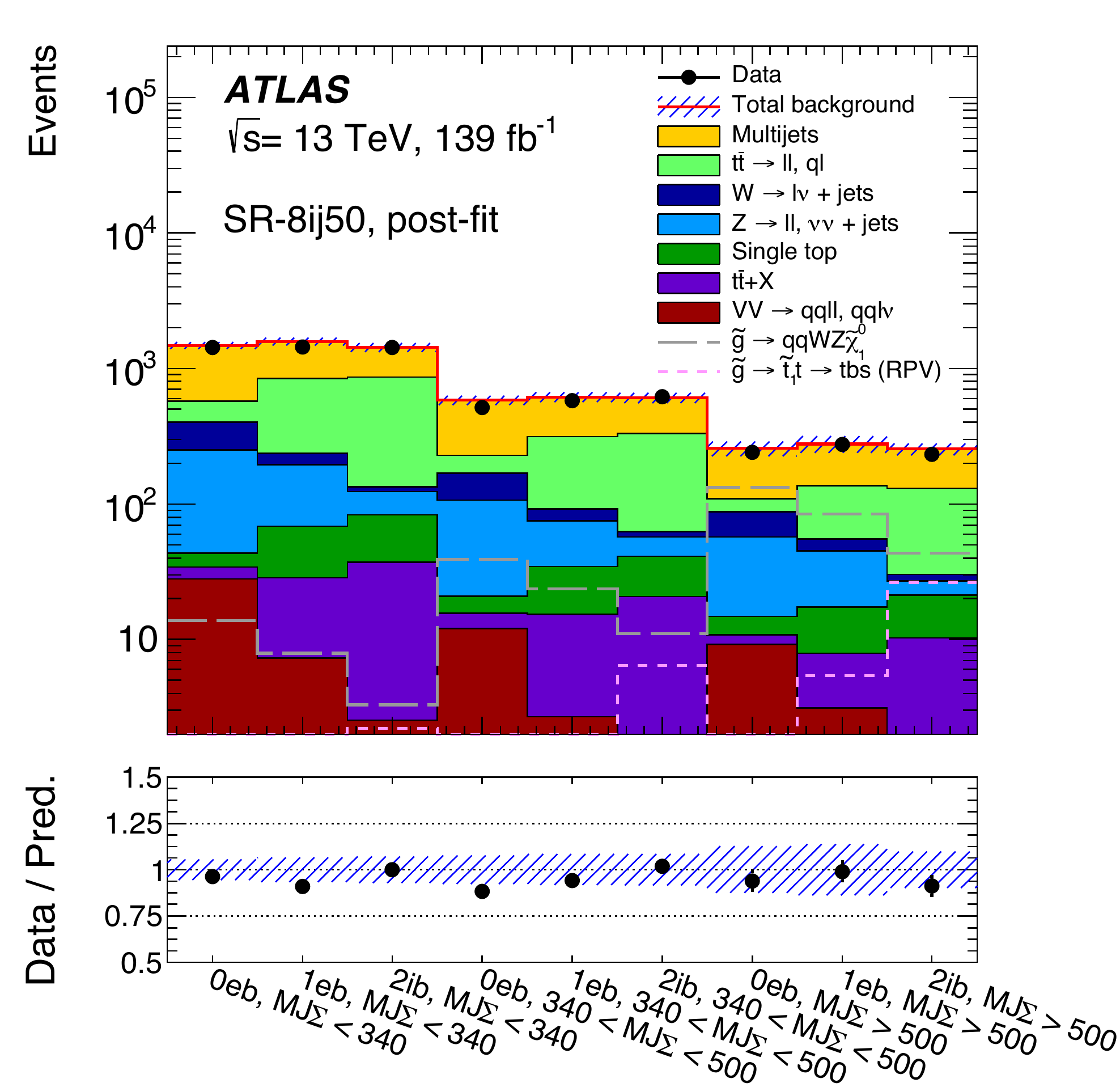}}
\hspace{0.5ex}
\subfloat[SR-9ij50, multi-bin]{\includegraphics[width=0.49\textwidth]{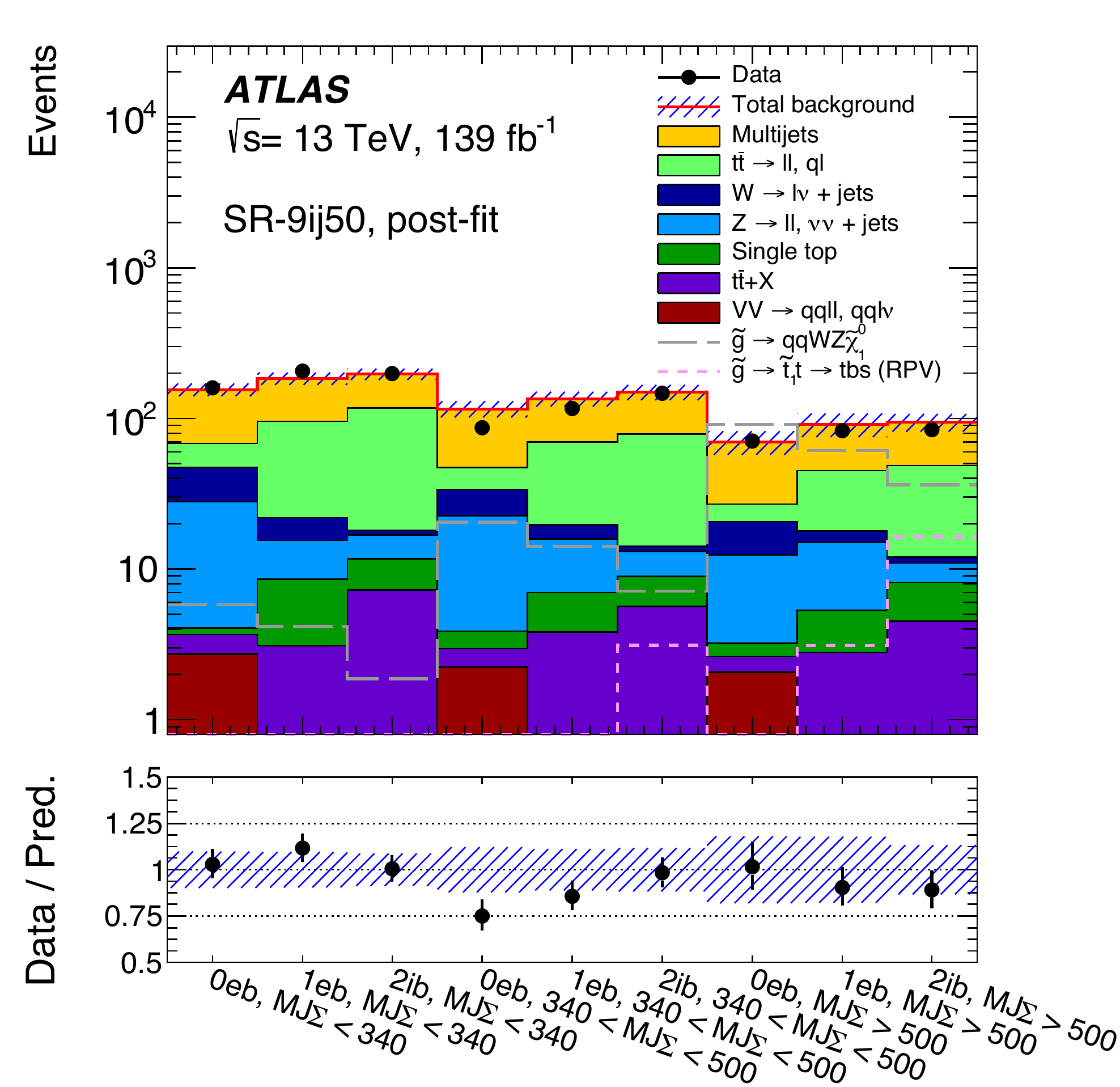}}
\\
\subfloat[SR-10ij50, multi-bin]{\includegraphics[width=0.49\textwidth]{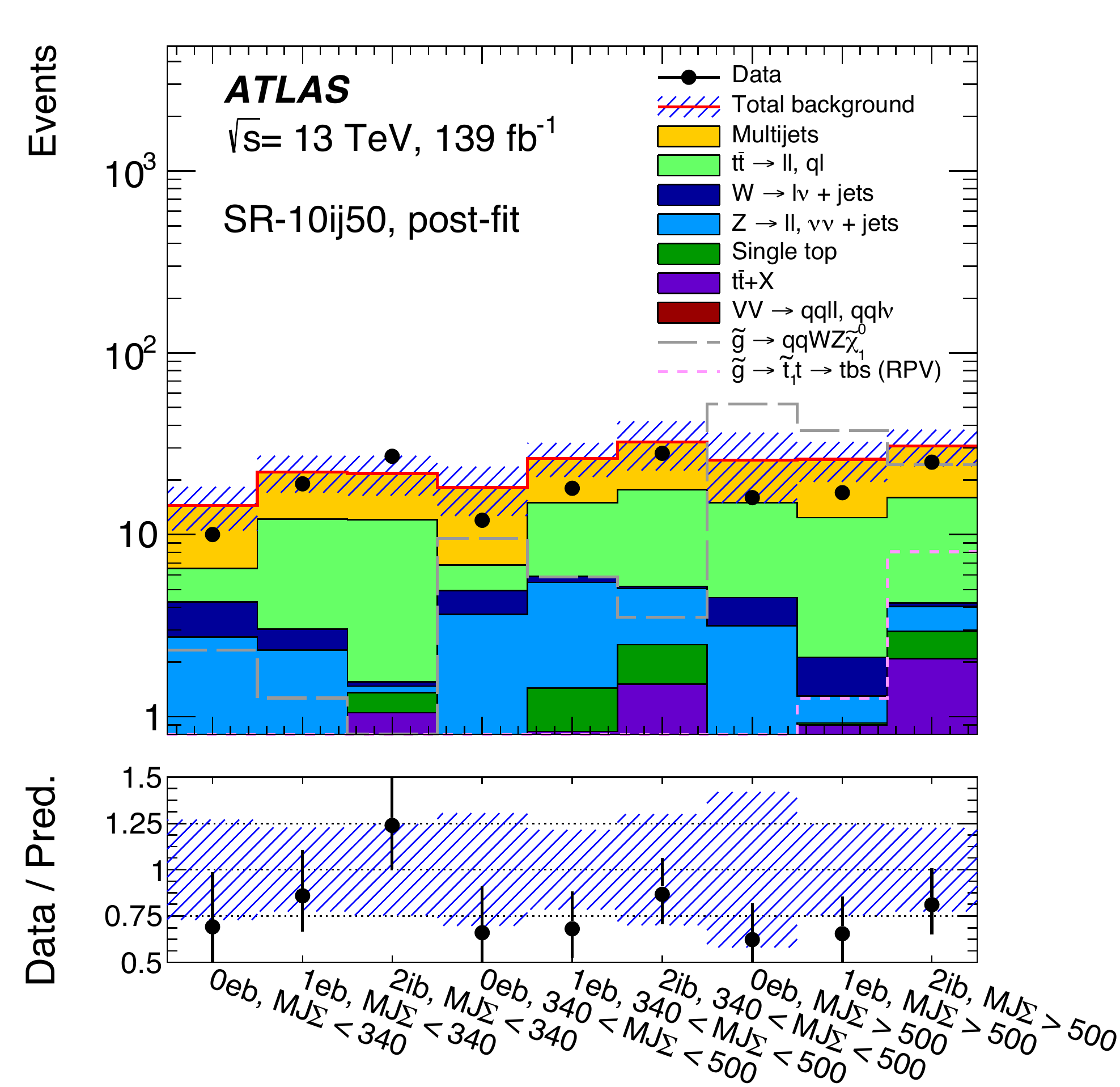}}
\hspace{0.5ex}
\subfloat[Single-bin summary]{\includegraphics[width=0.49\textwidth]{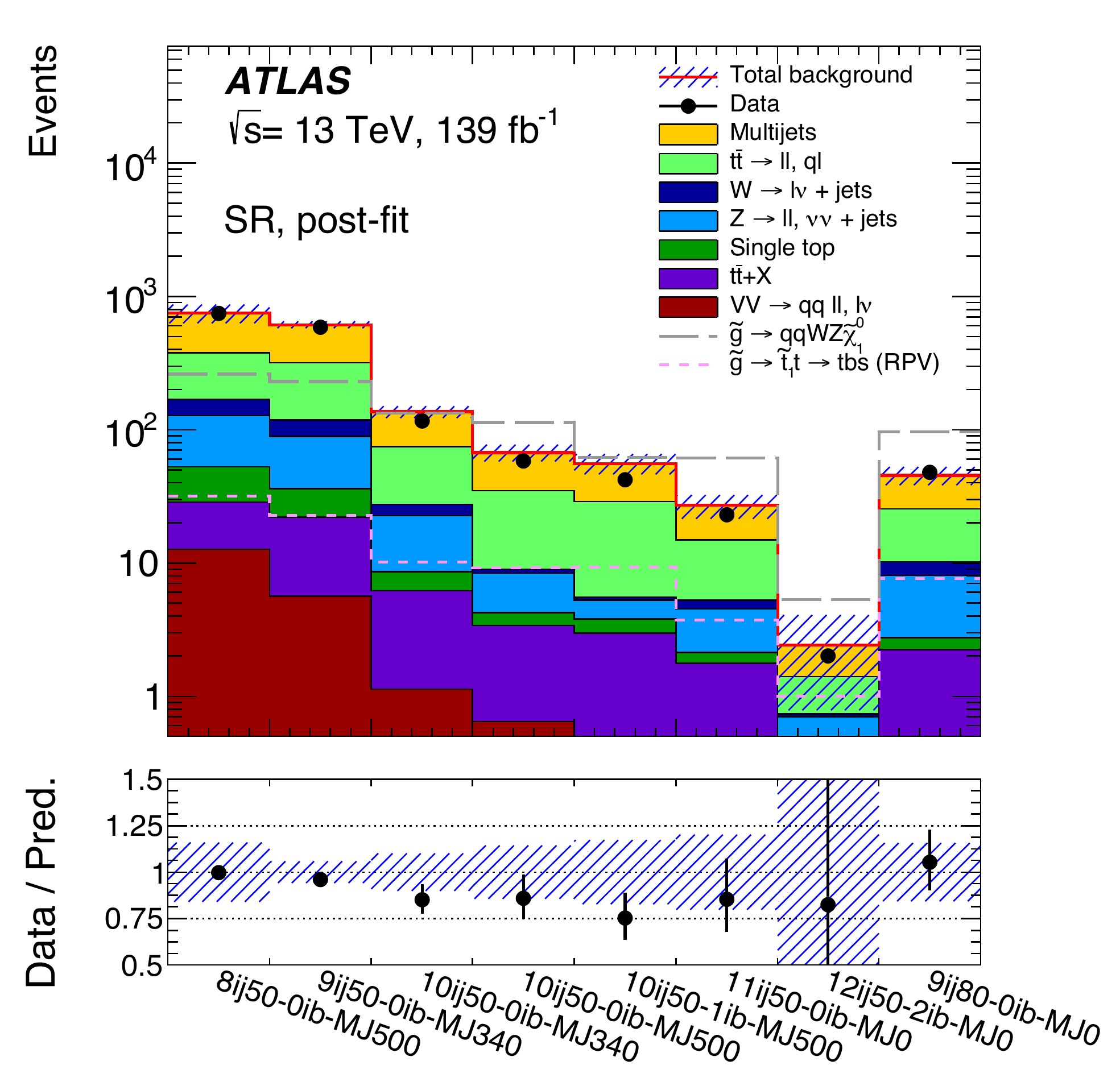}}
\caption{
Event yields in all signal regions, comparing data with the post-fit background predictions.
\dataMCcaption\,\signalcaption\,\ratiopanelcaption\,}
\label{fig:results:SRyields}
\end{figure}
 
\begin{table}
\centering
\renewcommand{\arraystretch}{1.1}
% The next lines are included from the .//tables/yields_singlebin_j89.tex input file
\begin{tabular}{@{\extracolsep{\fill}}lcc}
\toprule
Background process & \texttt{SR-8ij50-0ib-MJ500} & \texttt{SR-9ij50-0ib-MJ340} \\
\midrule
Multijets & $372 \pm 46$ & $296 \pm 39$ \\
Top quark pairs & $208 \pm 89$ & $201 \pm 65$ \\
\Wjets & ~~$41 \pm 20$ & ~~$29 \pm 12$ \\
\Zjets & ~~$75 \pm 46$ & ~~$53 \pm 33$ \\
Single top quarks & ~~$23.9 \pm 10.5$ & $14.1 \pm 5.3$ \\
$\ttbar + X$ & $16.1 \pm 8.3$ & $16.4 \pm 8.4$ \\
Diboson, triboson & $12.7 \pm 1.7$ & ~~$5.6 \pm 1.3$ \\
\midrule
Total background & ~~$750 \pm 120$ & $614 \pm 97$ \\
Observed events & $747$ & $588$ \\
\bottomrule
\end{tabular}
% End of text imported from the .//tables/yields_singlebin_j89.tex input file
\\[4ex]
% The next lines are included from the .//tables/yields_singlebin_j10.tex input file
\begin{tabular}{@{\extracolsep{\fill}}lccc}
\toprule
Background process & \texttt{SR-10ij50-0ib-MJ340} & \texttt{SR-10ij50-0ib-MJ500} & \texttt{SR-10ij50-1ib-MJ500} \\
\midrule
Multijets & $62 \pm 11$ & $32.8 \pm 6.1$~~ & $26.9 \pm 6.8$~~ \\
Top quark pairs & $47 \pm 17$ & $26 \pm 15$ & $23.5 \pm 9.8$~~ \\
\Wjets & $4.9 \pm 2.7$ & $0.6_{-0.6}^{+1.0}$~ & $0.3_{-0.3}^{+0.4}$~ \\
\Zjets & $14 \pm 13$ & $4 \pm 4$ & $1.5_{-1.5}^{+1.6}$~ \\
Single top quarks & $2.4 \pm 1.7$ & $0.9_{-0.9}^{+1.7}$~ & $0.8_{-0.8}^{+1.1}$~ \\
$\ttbar + X$ & $5.0 \pm 2.6$ & $2.7 \pm 1.5$ & $2.7 \pm 1.5$ \\
Diboson, triboson & $1.13 \pm 0.24$ & $0.65 \pm 0.19$ & $0.26 \pm 0.06$ \\
\midrule
Total background & $136 \pm 28$~~ & $68 \pm 18$ & $56 \pm 13$ \\
Observed events & $116$~~ & $58$ & $42$ \\
\bottomrule
\end{tabular}
% End of text imported from the .//tables/yields_singlebin_j10.tex input file
\\[4ex]
% The next lines are included from the .//tables/yields_singlebin_jmax.tex input file
\begin{tabular}{@{\extracolsep{\fill}}lccc}
\toprule
Background process & \texttt{SR-11ij50} & \texttt{SR-12ij50-2ib} & \texttt{SR-9ij80} \\
\midrule
Multijets & $12.0 \pm 3.8$~~ & $1.03 \pm 0.99$ & $20.0 \pm 5.1$ \\
Top quark pairs & $9.7 \pm 4.5$ & $0.66 \pm 0.53$ & $15.3 \pm 4.0$ \\
\Wjets & $0.75 \pm 0.75$ & $0.04_{-0.04}^{+0.08}$~\, & ~~$2.2 \pm 1.2$ \\
\Zjets & $2.4 \pm 1.4$ & $0.50 \pm 0.31$ & ~$5.3_{-5.3}^{+5.4}$ \\
Single top quarks & $0.4_{-0.4}^{+1.6}$~\, & $0.00 \pm 0.00$ & $0.50_{-0.50}^{+0.71}$ \\
$\ttbar + X$ & $1.4 \pm 0.8$ & $0.20 \pm 0.19$ & ~~$1.9 \pm 1.0$ \\
Diboson, triboson & $0.31 \pm 0.06$ & $0.00 \pm 0.00$ & ~~$0.32 \pm 0.06$ \\
\midrule
Total background & $27.0 \pm 7.3$~~ & $2.4 \pm 1.7$ & $45.6 \pm 9.5$ \\
Observed events & $23$ & $2$ & $48$ \\
\bottomrule
\end{tabular}
% End of text imported from the .//tables/yields_singlebin_jmax.tex input file
 
\caption{
Detailed event yields in all single-bin signal regions.
Due to correlations, summing the uncertainties of individual background components may not reproduce the final uncertainty of the total background prediction.
}
\label{tab:results:singlebinyields}
\end{table}

\begin{figure}[htbp]
\centering
\subfloat[SR-10ij50-0ib-MJ340]{\includegraphics[width=0.49\textwidth]{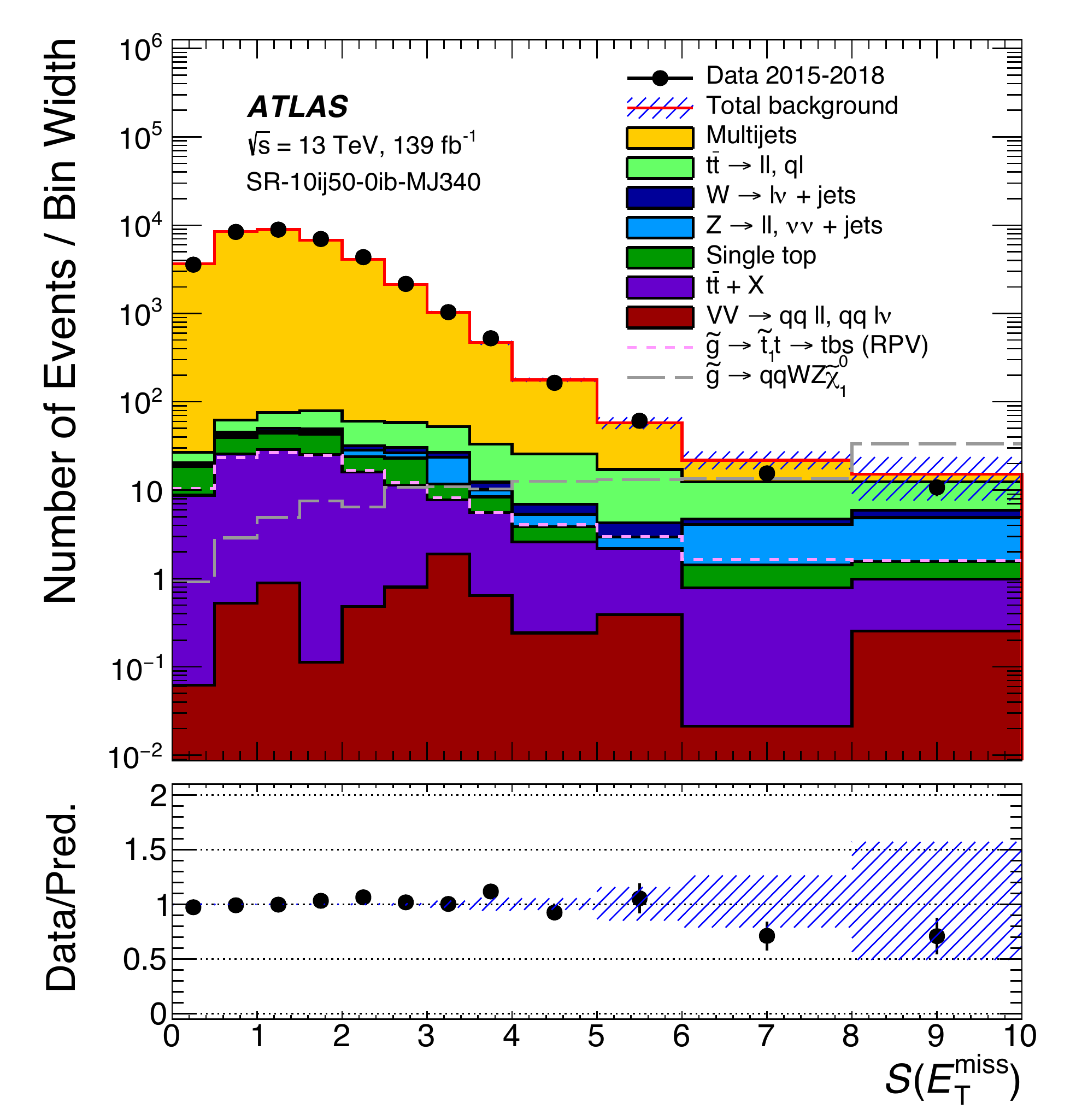}}
\hspace{0.5ex}
\subfloat[SR-12ij50-2ib]{\includegraphics[width=0.49\textwidth]{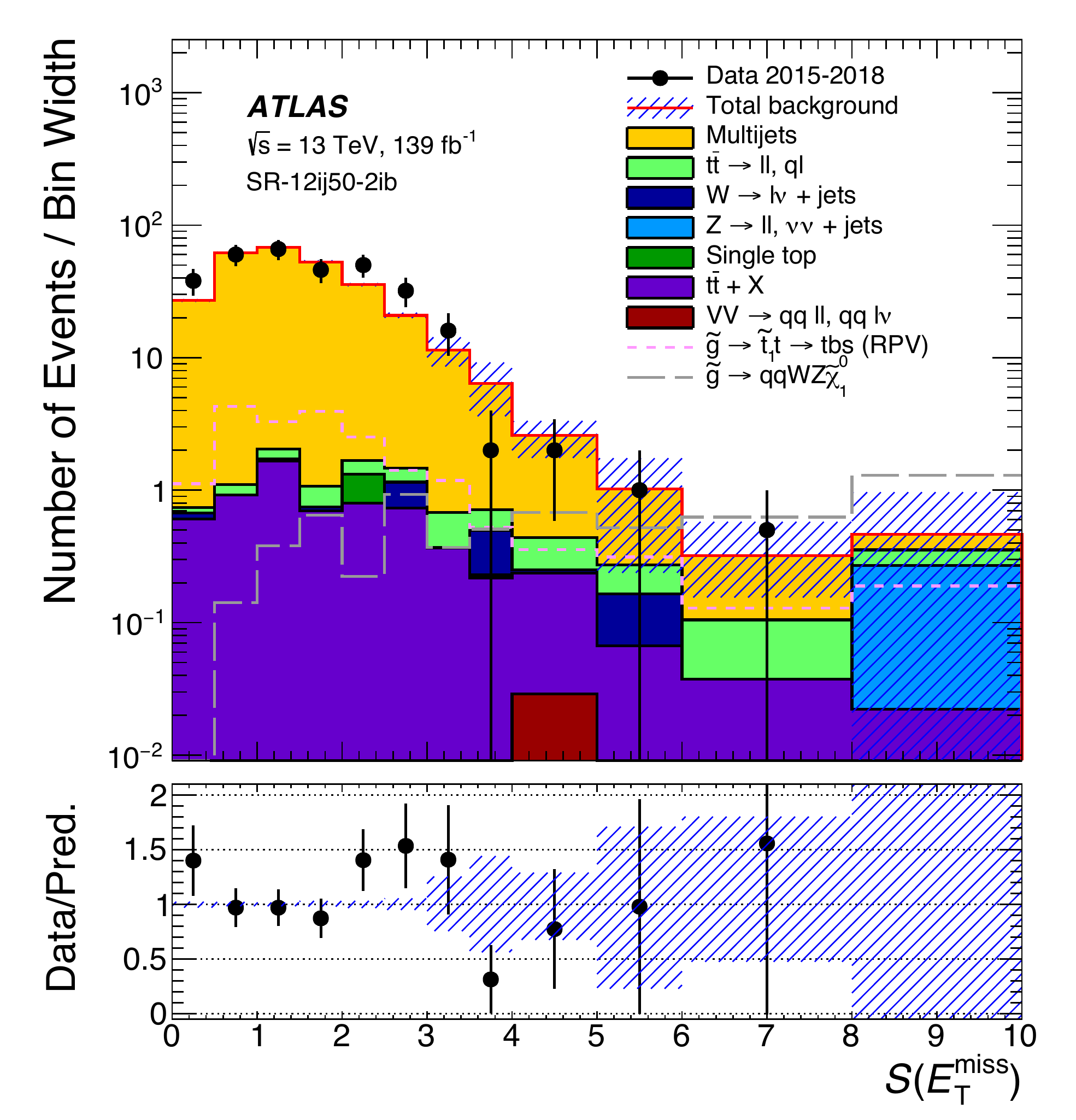}}
\\
\subfloat[SR-9ij80]{\includegraphics[width=0.49\textwidth]{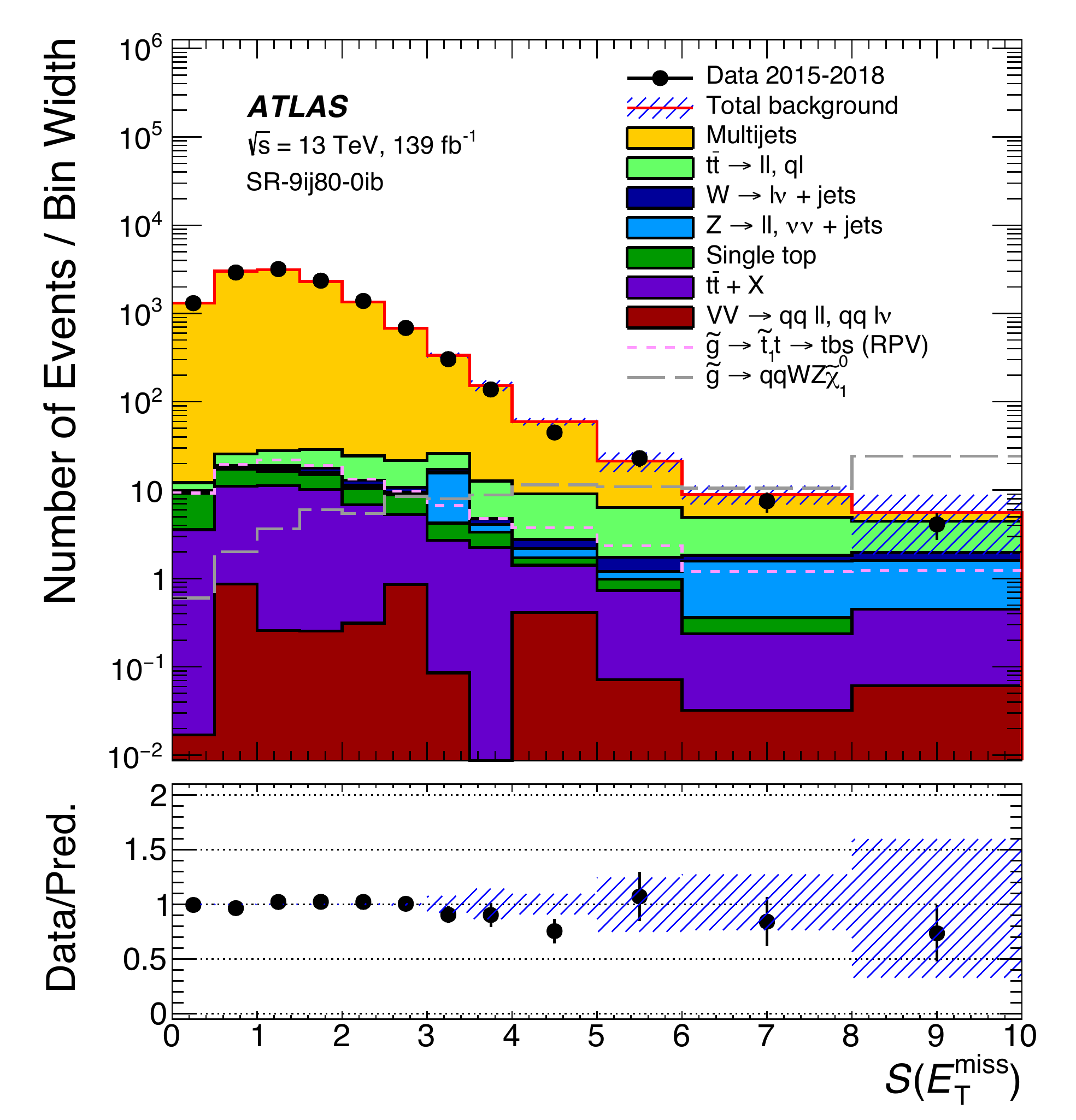}}
\caption{
Distribution of \metsig in selected signal regions.
The \ttbar and \Wjets backgrounds are normalised according to the background fits, and the multijet background prediction adjusted accordingly.
\dataMCcaption\,\signalcaption\,\ratiopanelcaption\,}
\label{fig:results:SRmetsig}
\end{figure}

The impact of the particle-flow reconstruction was assessed in the single-bin signal regions by applying the same event selection but instead using calorimeter-based hadronic reconstruction.
As a result of the improved jet and \met\ resolution, the multijet background is found to be reduced by 30--50\%, corresponding to two standard deviations when considering statistical uncertainties as well as the systematic uncertainties of the jet energy scale and multijet estimate closure, resulting in an overall 20\% lower total background yield in most regions.
Consequently, the expected sensitivity to new physics signals is improved by up to 30\% (one standard deviation), quantified in terms of the upper limit on BSM events.
% End of text imported from the .//source/results.tex input file

% The next lines are included from the .//source/interpretation.tex input file
\label{sec:interpretation}
 
The single-bin signal region event yields are used to derive model-independent constraints on the production of BSM particles.
\Tab{\ref{tab:results:upperlimits}} shows the observed 95\% CL limits on the visible cross-section $\langle\epsilon\sigma\rangle^{95}_\mathrm{obs}$ as well as the observed (expected) limits on the number of BSM signal events $S^{95}_\mathrm{obs}$ ($S^{95}_\mathrm{exp}$) in each signal region.
The discovery $p$-value $p(s=0)$, defined as the probability of observing at least the observed event yield when assuming that no signal is present, is calculated, as is the corresponding Gaussian significance $Z$.
For signal regions in which the predictions exceed the data, the value of $p(s=0)$ is capped at 0.5.
The smallest background $p$-value $\mathrm{CL}_\mathrm{b}$ computed in any region is 0.16, while the smallest discovery $p$-value is 0.41, and hence all observations are compatible with the SM-only hypothesis.
The most stringent limits observed are for the \texttt{SR-12ij50-2ib} selection, for which visible cross-sections greater than 40~ab are excluded.
 
% The next lines are included from the .//tables/ul_model_independent.tex input file
\begin{table}[htbp]
\centering
\renewcommand\arraystretch{1.1}
\setlength{\tabcolsep}{0.0pc}
\begin{tabular*}{\textwidth}{@{\extracolsep{\fill}}lccccc}
\noalign{\smallskip}\hline\noalign{\smallskip}
{Signal channel}                        & $\langle\epsilon{\mathrm \sigma}\rangle_\mathrm {obs}^{95}$~[fb]  &  $S_\mathrm{obs}^{95}$  & $S_\mathrm{exp}^{95}$ & $\mathrm{CL}_\mathrm{b}$ & $p(s=0)$ ($Z$)  \\
\noalign{\smallskip}\hline\noalign{\smallskip}
\texttt{SR-8ij50-0ib-MJ500}    & $1.16$ &  $163$~~ & $ { 162 }^{ +33 }_{ -39 }$ & $0.50$ & $ 0.50$~$(0.00)$ \\%
\texttt{SR-9ij50-0ib-MJ340}    & $0.95$ &  $133$~~ & $ { 140 }^{ +30 }_{ -31 }$ & $0.50$ & $ 0.50$~$(0.00)$ \\%
\texttt{SR-10ij50-0ib-MJ340}    & $0.22$ &  $31$ & ~~$ { 40 }^{ +15 }_{ -11 }$ & $0.21$ & $ 0.50$~$(0.00)$ \\%
\texttt{SR-10ij50-0ib-MJ500}    & $0.16$ &  ~~~$21.9$ & $ { 25.9 }^{ +9.6 }_{ -6.8 }$ & $0.29$ & $ 0.50$~$(0.00)$ \\%
\texttt{SR-10ij50-1ib-MJ500}    & $0.12$ &  ~~~$16.8$ & $ { 22.8 }^{ +8.5 }_{ -6.1 }$ & $0.16$ & $ 0.50$~$(0.00)$ \\%
\texttt{SR-11ij50}    & $0.09$ &  ~~~$13.0$ & $ { 15.1 }^{ +6.0 }_{ -4.2 }$ & $0.32$ & $ 0.50$~$(0.00)$ \\%
\texttt{SR-12ij50-2ib}    & $0.04$ &  ~~~~~$5.0$ & ~~$ { 5.2 }^{ +2.5 }_{ -1.7 }$ & $0.44$ & $ 0.50$~$(0.00)$ \\%
\texttt{SR-9ij80}    & $0.18$ &  ~~~$25.2$ & $ { 24.5 }^{ +7.0 }_{ -6.3 }$ & $0.57$ & $ 0.41$~$(0.22)$ \\%
\noalign{\smallskip}\hline\noalign{\smallskip}
\end{tabular*}
\caption[Breakdown of upper limits.]{
Left to right: 95\% CL upper limits on the visible cross-section ($\langle\epsilon\sigma\rangle_\mathrm{obs}^{95}$) and on the number of signal events ($S_\mathrm{obs}^{95}$ ).
The third column
($S_\mathrm{exp}^{95}$) shows the 95\% CL upper limit on the number of signal events, given the expected number of background events (and the $\pm 1\sigma$
excursions around the expectation).
The last two columns indicate the $\mathrm{CL}_\mathrm{b}$ value, i.e. the confidence level observed for the background-only hypothesis, and the discovery $p$-value ($p(s = 0)$), together with the corresponding Gaussian significance ($Z$).
}
\label{tab:results:upperlimits}
\end{table}
% End of text imported from the .//tables/ul_model_independent.tex input file
 
Constraints on sparticle production in several benchmark parameter planes are shown in \Fig{\ref{fig:interp:limits}}.
These limits extend beyond those achieved by the previous search~\cite{SUSY-2016-13}.
All exclusion fits are consistent with the background-only hypothesis, considering the total uncertainties.
 
For all three signal scenarios, the multi-bin limits provide the strongest expected constraints.
There are two exceptions: the \texttt{SR-9ij80-0ib} selection reaches parity with the multi-bin limits for small neutralino masses in the two-step gluino decay model, while at low gluino masses in the RPV decay the \texttt{SR-8ij50-0ib-MJ500} single-bin region achieves better sensitivity for small mass-splittings, but only for signal models already excluded by previous searches.
More significantly, the multi-bin approach enhances the sensitivity to RPV decays of the gluino, as well as to gluino-mediated stop production in the region where the LSP is moderately massive.
The 8-jet multi-bin region is most sensitive where the gluino--neutralino mass-splitting is smaller, while the 9- and 10-jet selections take over as the mass-splitting increases, which allows better signal--background discrimination with a harsher selection.
 
The strongest bounds on the gluino mass are placed at $\mgl=2~\TeV$ for gluinos decaying via a two-step cascade to $W$ and $Z$ bosons when assuming nearly massless neutralinos.
For gluino decays into a \ttbar pair and \met via RPC couplings, the region with $\mgl<1.8~\TeV$ and $\mlsp<700~\GeV$ is excluded, with the highest excluded neutralino mass being $\mlsp\simeq950~\GeV$.
This represents an improvement of $260~\GeV$ in \mgl and $230~\GeV$ in \mlsp with respect to the previous search results.
In the case of stop-mediated gluino decays via RPV couplings, the reach in \mgl is above $1.5~\TeV$ for \stopone masses of $400~\GeV$ to $1.1~\TeV$, with the maximal reach being achieved almost $1.6~\TeV$ for $m_{\stopone} = 900~\GeV$.
 
\begin{figure}
\centering
\subfloat[$\tilde{g}$, 2-step cascade]{\includegraphics[width=0.49\textwidth,clip,trim=45pt 0 0 0]{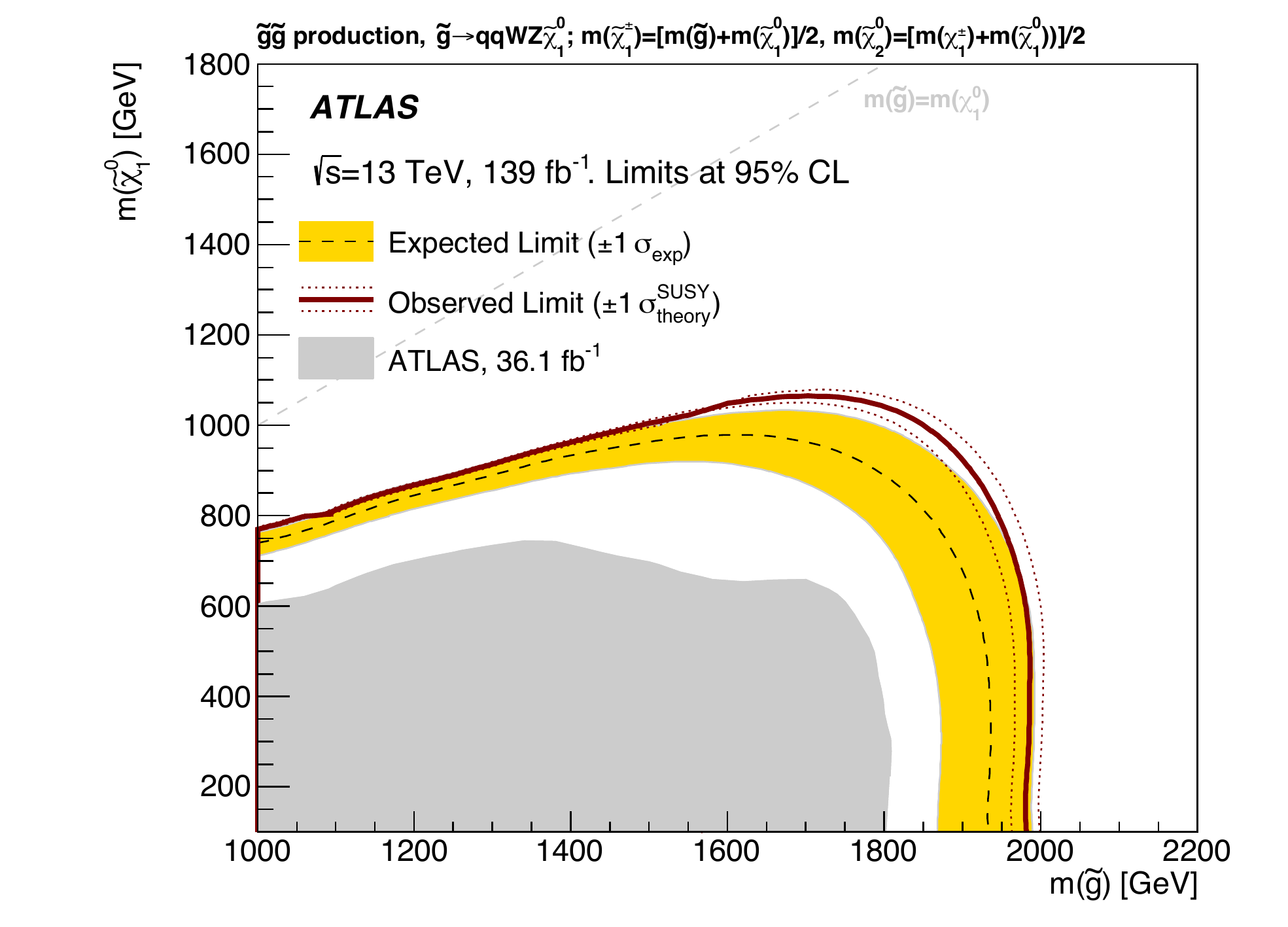}}
\hspace{0.5ex}
\subfloat[$\tilde{g} \rightarrow t \bar{t} \tilde{\chi}_1^0$]{\includegraphics[width=0.49\textwidth,clip,trim=45pt 0 0 0]{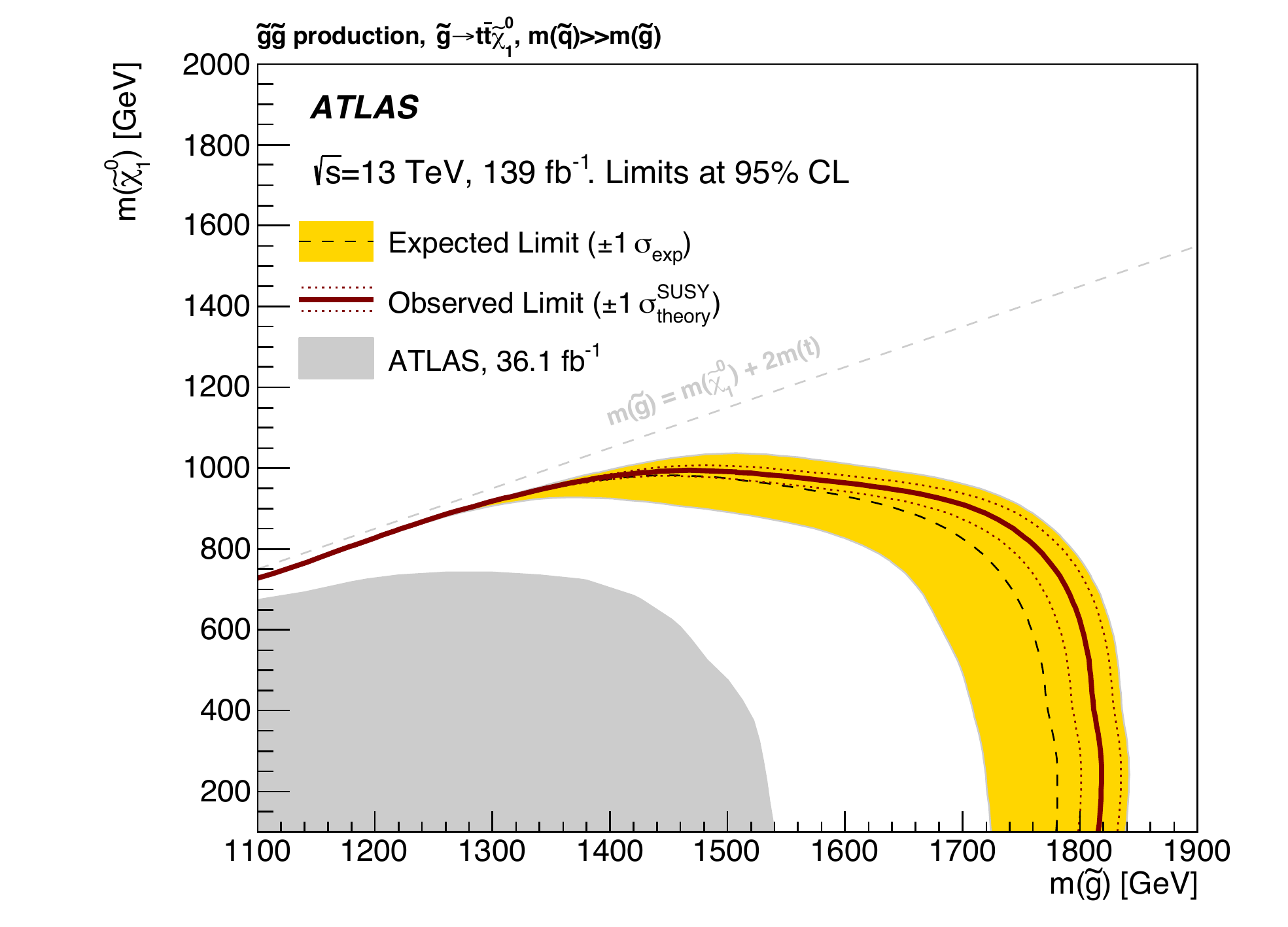}}
\\
\subfloat[RPV $\tilde{g} \rightarrow tbd,tbs$]{\includegraphics[width=0.49\textwidth,clip,trim=45pt 0 0 0]{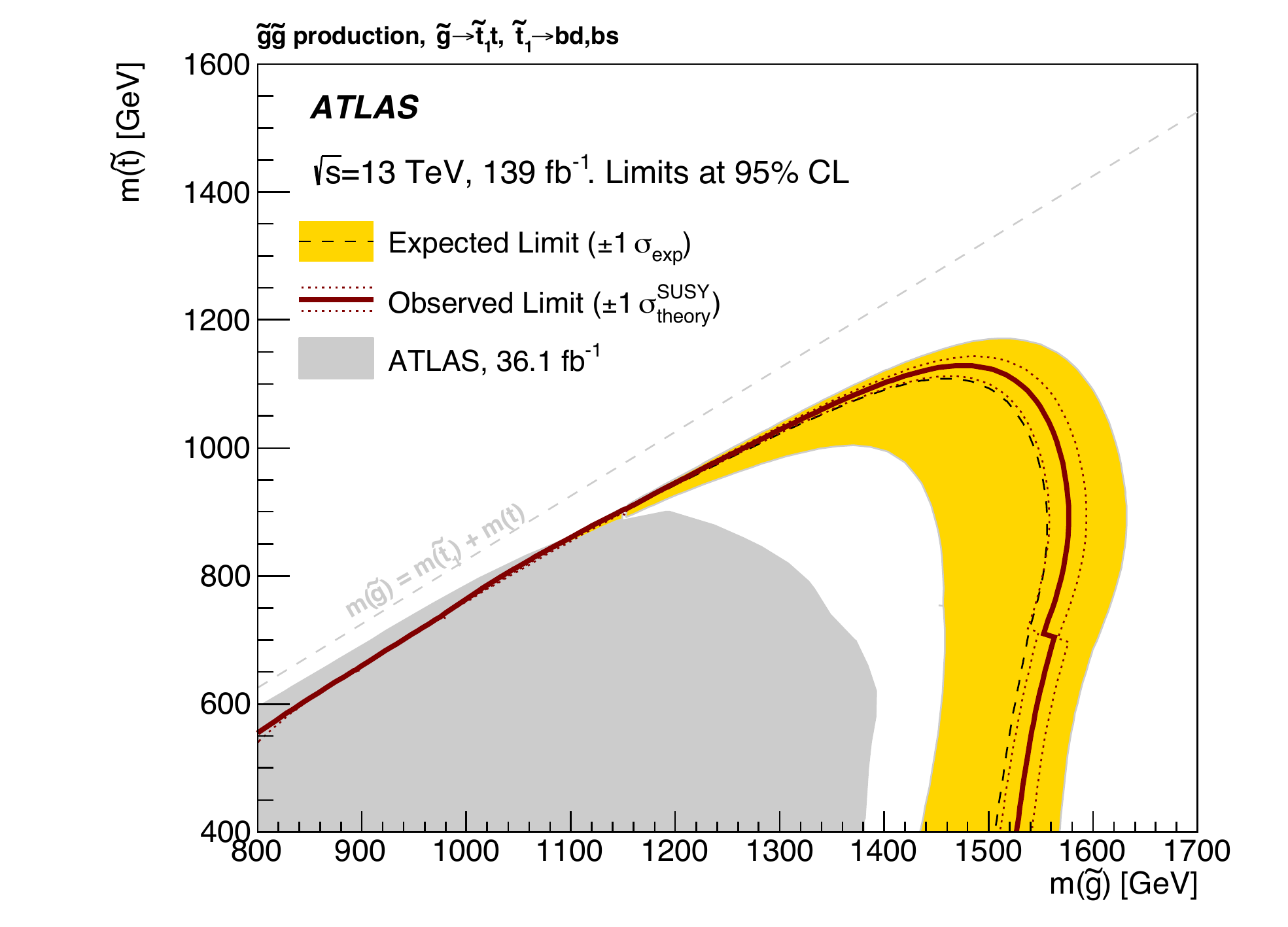}}
\caption{
Exclusion limits on gluino production in various parameter planes.
The solid maroon line indicates the observed exclusion limit at 95\% CL.
Transitions between two signal regions which provide the best expected sensitivity in different regions of the parameter space may lead to discontinuities in the observed limit.
The $1\sigma$ variations in the observed limit due to the uncertainty on the signal production cross-sections are shown with dotted maroon lines.
A dashed black line and yellow band respectively indicate the expected limit and its $1\sigma$ variations due to all uncertainties in the signal acceptance and background yields.
Grey shading is used to indicate the observed limit from the previous publication~\cite{SUSY-2016-13}.
}
\label{fig:interp:limits}
\end{figure}

% End of text imported from the .//source/interpretation.tex input file

\FloatBarrier
 
% The next lines are included from the .//source/conclusion.tex input file
\section{Conclusion}
\label{sec:conclusion}
 
A search for new physics, producing many jets and moderate \met, is presented, using the \intlumi~\ifb\ dataset of 13~\TeV\ $pp$ collisions collected by ATLAS during Run 2 of the LHC.
The analysis selects events in regions with large jet multiplicities (from $\geq$8 to $\geq$12 jets) with further requirements on the number of $b$-tagged jets, and on the sum of the masses of large-radius jets.
This search is the first from ATLAS to exploit particle-flow reconstruction for jets and \met.
The combination of this improvement together with a better object-based \metsig definition and a multi-bin statistical analysis leads to significant improvements in the search sensitivity beyond that afforded simply by the larger dataset.
 
No significant deviations from the Standard Model expectations are observed, while limits on the production of new particles are significantly extended.
In the context of $R$-parity conserving supersymmetry, limits on the gluino mass reach $2\,\TeV$ in the case of a two-step cascade via vector bosons, which are the most stringent limits observed to date when assuming a low-mass $\ninoone$.
For gluino-pair production followed by the decay $\gluino\to t \bar{t} \ninoone$ gluino masses up to $1.8~\TeV$ are excluded for  $\ninoone$ masses up to $700~\GeV$.
In the case of gluino-pair production followed by the $R$-parity-violating decays, $\gluino\to tbd,tbs$, limits extend up to almost $1.6~\TeV$ for a $\tilde t$ mass of $900~\GeV$.
Model-independent upper limits on the visible cross-section are set as low as 40~ab, in signal regions that probe extremely large jet multiplicities, with the most stringent selections requiring $\geq$12 jets of which two are $b$-tagged.

% End of text imported from the .//source/conclusion.tex input file
 
\section*{Acknowledgements}
 
% The next lines are included from the .//acknowledgements/Acknowledgements.tex input file

We thank CERN for the very successful operation of the LHC, as well as the
support staff from our institutions without whom ATLAS could not be
operated efficiently.
 
We acknowledge the support of ANPCyT, Argentina; YerPhI, Armenia; ARC, Australia; BMWFW and FWF, Austria; ANAS, Azerbaijan; SSTC, Belarus; CNPq and FAPESP, Brazil; NSERC, NRC and CFI, Canada; CERN; CONICYT, Chile; CAS, MOST and NSFC, China; COLCIENCIAS, Colombia; MSMT CR, MPO CR and VSC CR, Czech Republic; DNRF and DNSRC, Denmark; IN2P3-CNRS and CEA-DRF/IRFU, France; SRNSFG, Georgia; BMBF, HGF and MPG, Germany; GSRT, Greece; RGC and Hong Kong SAR, China; ISF and Benoziyo Center, Israel; INFN, Italy; MEXT and JSPS, Japan; CNRST, Morocco; NWO, Netherlands; RCN, Norway; MNiSW and NCN, Poland; FCT, Portugal; MNE/IFA, Romania; MES of Russia and NRC KI, Russia Federation; JINR; MESTD, Serbia; MSSR, Slovakia; ARRS and MIZ\v{S}, Slovenia; DST/NRF, South Africa; MICINN, Spain; SRC and Wallenberg Foundation, Sweden; SERI, SNSF and Cantons of Bern and Geneva, Switzerland; MOST, Taiwan; TAEK, Turkey; STFC, United Kingdom; DOE and NSF, United States of America. In addition, individual groups and members have received support from BCKDF, CANARIE, Compute Canada and CRC, Canada; ERC, ERDF, Horizon 2020, Marie Sk{\l}odowska-Curie Actions and COST, European Union; Investissements d'Avenir Labex, Investissements d'Avenir Idex and ANR, France; DFG and AvH Foundation, Germany; Herakleitos, Thales and Aristeia programmes co-financed by EU-ESF and the Greek NSRF, Greece; BSF-NSF and GIF, Israel; La Caixa Banking Foundation, CERCA Programme Generalitat de Catalunya and PROMETEO and GenT Programmes Generalitat Valenciana, Spain; G\"{o}ran Gustafssons Stiftelse, Sweden; The Royal Society and Leverhulme Trust, United Kingdom.
 
The crucial computing support from all WLCG partners is acknowledged gratefully, in particular from CERN, the ATLAS Tier-1 facilities at TRIUMF (Canada), NDGF (Denmark, Norway, Sweden), CC-IN2P3 (France), KIT/GridKA (Germany), INFN-CNAF (Italy), NL-T1 (Netherlands), PIC (Spain), ASGC (Taiwan), RAL (UK) and BNL (USA), the Tier-2 facilities worldwide and large non-WLCG resource providers. Major contributors of computing resources are listed in Ref.~\cite{ATL-SOFT-PUB-2020-001}.
 
% End of text imported from the .//acknowledgements/Acknowledgements.tex input file

\printbibliography
 
\clearpage% ATLAS Collaboration author list
% Reference date of SUSY-2018-17 is 2020-01-24
% Author list last updated on date 13-AUG-20
% Data extracted on 13-Aug-2020 for paper reference SUSY-2018-17
% at 7:10pm
 
\begin{flushleft}
\hypersetup{urlcolor=black}
{\Large The ATLAS Collaboration}

\bigskip

\AtlasOrcid[0000-0002-6665-4934]{G.~Aad}$^\textrm{\scriptsize 102}$,    
\AtlasOrcid[0000-0002-5888-2734]{B.~Abbott}$^\textrm{\scriptsize 128}$,    
\AtlasOrcid{D.C.~Abbott}$^\textrm{\scriptsize 103}$,    
\AtlasOrcid[0000-0002-2788-3822]{A.~Abed~Abud}$^\textrm{\scriptsize 36}$,    
\AtlasOrcid[0000-0002-1002-1652]{K.~Abeling}$^\textrm{\scriptsize 53}$,    
\AtlasOrcid[0000-0002-2987-4006]{D.K.~Abhayasinghe}$^\textrm{\scriptsize 94}$,    
\AtlasOrcid[0000-0002-8496-9294]{S.H.~Abidi}$^\textrm{\scriptsize 166}$,    
\AtlasOrcid[0000-0002-8279-9324]{O.S.~AbouZeid}$^\textrm{\scriptsize 40}$,    
\AtlasOrcid{N.L.~Abraham}$^\textrm{\scriptsize 155}$,    
\AtlasOrcid[0000-0001-5329-6640]{H.~Abramowicz}$^\textrm{\scriptsize 160}$
% ...
\bigskip
\\

% ...
$^{36}$CERN, Geneva; Switzerland.\\
% ...
$^{40}$Niels Bohr Institute, University of Copenhagen, Copenhagen; Denmark.\\
% ...
$^{53}$II. Physikalisches Institut, Georg-August-Universit\"{a}t G\"ottingen, G\"ottingen; Germany.\\
% ...
$^{94}$Department of Physics, Royal Holloway University of London, Egham; United Kingdom.\\
% ...
$^{102}$CPPM, Aix-Marseille Universit\'e, CNRS/IN2P3, Marseille; France.\\
$^{103}$Department of Physics, University of Massachusetts, Amherst MA; United States of America.\\
% ...
$^{128}$Homer L. Dodge Department of Physics and Astronomy, University of Oklahoma, Norman OK; United States of America.\\
% ...
$^{155}$Department of Physics and Astronomy, University of Sussex, Brighton; United Kingdom.\\
% ...
$^{160}$Raymond and Beverly Sackler School of Physics and Astronomy, Tel Aviv University, Tel Aviv; Israel.\\
% ...
$^{166}$Department of Physics, University of Toronto, Toronto ON; Canada.\\
% ...

\end{flushleft}

% Created with Glance <Atlas.Glance@cern.ch>

\end{document}